\documentstyle[10pt,epsfig,dp_delphititle,amssymb,epsfig,cite,subfigure,amsmath,mathcomp]{dp_delphi}
\makeindex
\def\ifmath#1{\relax\ifmmode #1\else $#1$\fi}%
\newcommand{\bc}{{\begin{center}}}
\newcommand{\ec}{{\end{center}}}
\newcommand{\br}{{\begin{right}}}
\newcommand{\er}{{\end{right}}}
\newcommand{\beq}{\begin{equation}}
\newcommand{\eeq}{\end{equation}}
\newcommand{\beqn}{\begin{eqnarray}}
\newcommand{\eeqn}{\end{eqnarray}}
\newcommand{\ba}{\begin{array}{c}}
\newcommand{\bat}{\begin{array}{cc}}
\newcommand{\ea}{\end{array}}

\newcommand{\bi}{\begin{itemize}}
\newcommand{\ei}{\end{itemize}}

\newcommand{\ra}{{\rightarrow}}

\newcommand{\delphi}{{\sc Delphi}}
\newcommand{\dopal}{{\sc Opal}}

\newcommand{\daleph}{{\sc Aleph}}
\newcommand{\delsim}{{\sc Delsim}}

\newcommand{\Pythiav}{{\sc Pythia 6.156}}
\newcommand{\Herwigv}{{\sc Herwig 6.2}}
\newcommand{\Ariadnev}{{\sc Ariadne 4.08}}
\newcommand{\Pythia}{{\sc Pythia}}
\newcommand{\Herwig}{{\sc Herwig}}
\newcommand{\Ariadne}{{\sc Ariadne}}

\newcommand{\cambridge}{{\sc Cambridge}}

\newcommand{\lep}{{\sc Lep}}
\newcommand{\sld}{{\sc Sld}}
\newcommand{\slc}{{\sc Slc}}

\newcommand{\GeV}{\ifmath{\mathrm{GeV}}}

\newcommand{\GeVcTwo}{\ifmath{{\mathrm GeV}/c^2}}

\newcommand{\Order} {\ifmath{{\mathcal{O}}}}
\newcommand{\alphas} {\ifmath{\alpha_{s}}}

\newcommand{\MS} {\ifmath{\overline{MS}}}

\newcommand{\Mb}   {\ifmath{M_b}}

\newcommand{\MZ}   {\ifmath{M_Z}}
\newcommand{\mb}   {\ifmath{m_b}}

\newcommand{\Rb}   {\ifmath{R_b}}
\newcommand{\Rc}   {\ifmath{R_c}}

\newcommand{\Rfbl}  {\ifmath{R_4^{b\ell}}}

\newcommand{\Rthreebl}{\ifmath{R_{3}^{b\ell}}}
\newcommand{\Rtwobl}{\ifmath{R_{2}^{b\ell}}}

\newcommand{\Rnbl}  {\ifmath{R_n^{b\ell}}}

\newcommand{\Ztobb}  {\ifmath{Z\rightarrow b\bar{b}}}
\newcommand{\Ztoll}  {\ifmath{Z\rightarrow \ell\bar{\ell}}}

\newcommand{\qb}{\ifmath{b}}

\newcommand{\qq}{\ifmath{q\bar{q}}}
\newcommand{\bb}{\ifmath{b\bar{b}}}
\newcommand{\cc}{\ifmath{c\bar{c}}}

\newcommand{\gcc}{\ifmath{g_{c\overline{c}}}}
\newcommand{\gbb}{\ifmath{g_{b\overline{b}}}}
\newcommand{\ycut}{\ifmath{y_{cut}}}
\newcommand{\etal}{{\it et al}}

\def\DpPaperGroup{PH-EP}
\def\DpPaperRef{2007-011}
\def\DpDate{4 May 2007}
\def\DpAuthors{DELPHI Collaboration}
\def\DpSubmit{(Accepted by Eur. Phys. J. C )}
\def\DpTitle{{ Study of $b$-quark mass effects 
\\in multijet topologies
\\with the DELPHI detector at LEP
}}
\def\DpComment{}
\def\DpEMail{}

\begin{document}
\makeatletter
\makeatother

\begin{titlepage}
\pagenumbering{roman}

\CERNpreprint{\DpPaperGroup}{\DpPaperRef}   
\date{{\small\DpDate}}                      
\title{\DpTitle}                            
\address{\DpAuthors}                        

\begin{shortabs}                            
\noindent
%
The effect of the heavy \qb-quark mass on the two, three and four-jet rates
is studied using \lep~data collected by the \delphi~experiment at the $Z$ peak
in 1994 and 1995.
The rates of \qb-quark jets and light quark jets ($\ell = uds$) in 
events with $n = 2$, 3, and 4 jets, together with 
the ratio of two and four-jet rates of \qb-quarks with respect to light-quarks, 
\Rnbl, have been measured with a double-tag technique using
the \cambridge~jet-clustering algorithm.  
A comparison between experimental results and theory (matrix element or 
Monte Carlo event generators such as {\sc Pythia}, {\sc Herwig} and {\sc Ariadne}) 
is done after the hadronisation phase. 

Using the four-jet observable \Rfbl, 
a measurement of the \qb-quark mass using massive leading-order calculations 
gives: 
\[ \small{
\nonumber
m_b(M_Z) = 3.76 \pm 0.32~({\rm stat}) \pm 0.17 ~({\rm syst})  
                \pm 0.22~({\rm had}) \pm 0.90~({\rm theo})~\GeVcTwo \; .} 
\] \noindent
This result is compatible with previous three-jet determinations at the $M_Z$
energy scale and with low energy mass measurements evolved to the $M_Z$ scale
using QCD Renormalisation Group Equations.  

\end{shortabs}

\vfill

\begin{center}
\DpSubmit \ \\          
\DpComment \ \\
\DpEMail \ \\
\end{center}

\vfill
\clearpage

\headsep 10.0pt

\addtolength{\textheight}{10mm}
\addtolength{\footskip}{-5mm}
\begingroup
%
\newcommand{\DpName}[2]{\hbox{#1$^{\ref{#2}}$},\hfill}
\newcommand{\DpNameTwo}[3]{\hbox{#1$^{\ref{#2},\ref{#3}}$},\hfill}
\newcommand{\DpNameThree}[4]{\hbox{#1$^{\ref{#2},\ref{#3},\ref{#4}}$},\hfill}
\newskip\Bigfill \Bigfill = 0pt plus 1000fill
\newcommand{\DpNameLast}[2]{\hbox{#1$^{\ref{#2}}$}\hspace{\Bigfill}}

%
\footnotesize
\noindent
\DpName{J.Abdallah}{LPNHE}
\DpName{P.Abreu}{LIP}
\DpName{W.Adam}{VIENNA}
\DpName{P.Adzic}{DEMOKRITOS}
\DpName{T.Albrecht}{KARLSRUHE}
\DpName{R.Alemany-Fernandez}{CERN}
\DpName{T.Allmendinger}{KARLSRUHE}
\DpName{P.P.Allport}{LIVERPOOL}
\DpName{U.Amaldi}{MILANO2}
\DpName{N.Amapane}{TORINO}
% ...
\normalsize
\endgroup
\newpage

\titlefoot{Department of Physics and Astronomy, Iowa State
     University, Ames IA 50011-3160, USA
    \label{AMES}}
\titlefoot{IIHE, ULB-VUB,
     Pleinlaan 2, B-1050 Brussels, Belgium
    \label{BRUSSELS}}
\titlefoot{Physics Laboratory, University of Athens, Solonos Str.
     104, GR-10680 Athens, Greece
    \label{ATHENS}}
\titlefoot{Department of Physics, University of Bergen,
     All\'egaten 55, NO-5007 Bergen, Norway
    \label{BERGEN}}
\titlefoot{Dipartimento di Fisica, Universit\`a di Bologna and INFN,
     Via Irnerio 46, IT-40126 Bologna, Italy
    \label{BOLOGNA}}
\titlefoot{Centro Brasileiro de Pesquisas F\'{\i}sicas, rua Xavier Sigaud 150,
     BR-22290 Rio de Janeiro, Brazil
    \label{BRASIL-CBPF}}
\titlefoot{Inst. de F\'{\i}sica, Univ. Estadual do Rio de Janeiro,
     rua S\~{a}o Francisco Xavier 524, Rio de Janeiro, Brazil
    \label{BRASIL-IFUERJ}}
\titlefoot{Coll\`ege de France, Lab. de Physique Corpusculaire, IN2P3-CNRS,
     FR-75231 Paris Cedex 05, France
    \label{CDF}}
\titlefoot{CERN, CH-1211 Geneva 23, Switzerland
    \label{CERN}}
\titlefoot{Institut de Recherches Subatomiques, IN2P3 - CNRS/ULP - BP20,
     FR-67037 Strasbourg Cedex, France
    \label{CRN}}
\titlefoot{Now at DESY-Zeuthen, Platanenallee 6, D-15735 Zeuthen, Germany
    \label{DESY}}
\titlefoot{Institute of Nuclear Physics, N.C.S.R. Demokritos,
     P.O. Box 60228, GR-15310 Athens, Greece
    \label{DEMOKRITOS}}
\titlefoot{FZU, Inst. of Phys. of the C.A.S. High Energy Physics Division,
     Na Slovance 2, CZ-182 21, Praha 8, Czech Republic
    \label{FZU}}
\titlefoot{Dipartimento di Fisica, Universit\`a di Genova and INFN,
     Via Dodecaneso 33, IT-16146 Genova, Italy
    \label{GENOVA}}
\titlefoot{Institut des Sciences Nucl\'eaires, IN2P3-CNRS, Universit\'e
     de Grenoble 1, FR-38026 Grenoble Cedex, France
    \label{GRENOBLE}}
\titlefoot{Helsinki Institute of Physics and Department of Physical Sciences,
     P.O. Box 64, FIN-00014 University of Helsinki, 
     \indent~~Finland
    \label{HELSINKI}}
\titlefoot{Joint Institute for Nuclear Research, Dubna, Head Post
     Office, P.O. Box 79, RU-101 000 Moscow, Russian Federation
    \label{JINR}}
\titlefoot{Institut f\"ur Experimentelle Kernphysik,
     Universit\"at Karlsruhe, Postfach 6980, DE-76128 Karlsruhe,
     Germany
    \label{KARLSRUHE}}
\titlefoot{Institute of Nuclear Physics PAN,Ul. Radzikowskiego 152,
     PL-31142 Krakow, Poland
    \label{KRAKOW1}}
\titlefoot{Faculty of Physics and Nuclear Techniques, University of Mining
     and Metallurgy, PL-30055 Krakow, Poland
    \label{KRAKOW2}}
\titlefoot{Universit\'e de Paris-Sud, Lab. de l'Acc\'el\'erateur
     Lin\'eaire, IN2P3-CNRS, B\^{a}t. 200, FR-91405 Orsay Cedex, France
    \label{LAL}}
\titlefoot{School of Physics and Chemistry, University of Lancaster,
     Lancaster LA1 4YB, UK
    \label{LANCASTER}}
\titlefoot{LIP, IST, FCUL - Av. Elias Garcia, 14-$1^{o}$,
     PT-1000 Lisboa Codex, Portugal
    \label{LIP}}
\titlefoot{Department of Physics, University of Liverpool, P.O.
     Box 147, Liverpool L69 3BX, UK
    \label{LIVERPOOL}}
\titlefoot{Dept. of Physics and Astronomy, Kelvin Building,
     University of Glasgow, Glasgow G12 8QQ, UK
    \label{GLASGOW}}
\titlefoot{LPNHE, IN2P3-CNRS, Univ.~Paris VI et VII, Tour 33 (RdC),
     4 place Jussieu, FR-75252 Paris Cedex 05, France
    \label{LPNHE}}
\titlefoot{Department of Physics, University of Lund,
     S\"olvegatan 14, SE-223 63 Lund, Sweden
    \label{LUND}}
\titlefoot{Universit\'e Claude Bernard de Lyon, IPNL, IN2P3-CNRS,
     FR-69622 Villeurbanne Cedex, France
    \label{LYON}}
\titlefoot{Dipartimento di Fisica, Universit\`a di Milano and INFN-MILANO,
     Via Celoria 16, IT-20133 Milan, Italy
    \label{MILANO}}
\titlefoot{Dipartimento di Fisica, Univ. di Milano-Bicocca and
     INFN-MILANO, Piazza della Scienza 3, IT-20126 Milan, Italy
    \label{MILANO2}}
\titlefoot{IPNP of MFF, Charles Univ., Areal MFF,
     V Holesovickach 2, CZ-180 00, Praha 8, Czech Republic
    \label{NC}}
\titlefoot{NIKHEF, Postbus 41882, NL-1009 DB
     Amsterdam, The Netherlands
    \label{NIKHEF}}
\titlefoot{National Technical University, Physics Department,
     Zografou Campus, GR-15773 Athens, Greece
    \label{NTU-ATHENS}}
\titlefoot{Physics Department, University of Oslo, Blindern,
     NO-0316 Oslo, Norway
    \label{OSLO}}
\titlefoot{Dpto. Fisica, Univ. Oviedo, Avda. Calvo Sotelo
     s/n, ES-33007 Oviedo, Spain
    \label{OVIEDO}}
\titlefoot{Department of Physics, University of Oxford,
     Keble Road, Oxford OX1 3RH, UK
    \label{OXFORD}}
\titlefoot{Dipartimento di Fisica, Universit\`a di Padova and
     INFN, Via Marzolo 8, IT-35131 Padua, Italy
    \label{PADOVA}}
\titlefoot{Rutherford Appleton Laboratory, Chilton, Didcot
     OX11 OQX, UK
    \label{RAL}}
\titlefoot{Dipartimento di Fisica, Universit\`a di Roma II and
     INFN, Tor Vergata, IT-00173 Rome, Italy
    \label{ROMA2}}
\titlefoot{Dipartimento di Fisica, Universit\`a di Roma III and
     INFN, Via della Vasca Navale 84, IT-00146 Rome, Italy
    \label{ROMA3}}
\titlefoot{DAPNIA/Service de Physique des Particules,
     CEA-Saclay, FR-91191 Gif-sur-Yvette Cedex, France
    \label{SACLAY}}
\titlefoot{Instituto de Fisica de Cantabria (CSIC-UC), Avda.
     los Castros s/n, ES-39006 Santander, Spain
    \label{SANTANDER}}
\titlefoot{Inst. for High Energy Physics, Serpukov
     P.O. Box 35, Protvino, (Moscow Region), Russian Federation
    \label{SERPUKHOV}}
\titlefoot{J. Stefan Institute, Jamova 39, SI-1000 Ljubljana, Slovenia
    \label{SLOVENIJA1}}
\titlefoot{Laboratory for Astroparticle Physics,
     University of Nova Gorica, Kostanjeviska 16a, SI-5000 Nova Gorica, Slovenia
    \label{SLOVENIJA2}}
\titlefoot{Department of Physics, University of Ljubljana,
     SI-1000 Ljubljana, Slovenia
    \label{SLOVENIJA3}}
\titlefoot{Fysikum, Stockholm University,
     Box 6730, SE-113 85 Stockholm, Sweden
    \label{STOCKHOLM}}
\titlefoot{Dipartimento di Fisica Sperimentale, Universit\`a di
     Torino and INFN, Via P. Giuria 1, IT-10125 Turin, Italy
    \label{TORINO}}
\titlefoot{INFN,Sezione di Torino and Dipartimento di Fisica Teorica,
     Universit\`a di Torino, Via Giuria 1,
     IT-10125 Turin, Italy
    \label{TORINOTH}}
\titlefoot{Dipartimento di Fisica, Universit\`a di Trieste and
     INFN, Via A. Valerio 2, IT-34127 Trieste, Italy
    \label{TRIESTE}}
\titlefoot{Istituto di Fisica, Universit\`a di Udine and INFN,
     IT-33100 Udine, Italy
    \label{UDINE}}
\titlefoot{Univ. Federal do Rio de Janeiro, C.P. 68528
     Cidade Univ., Ilha do Fund\~ao
     BR-21945-970 Rio de Janeiro, Brazil
    \label{UFRJ}}
\titlefoot{Department of Radiation Sciences, University of
     Uppsala, P.O. Box 535, SE-751 21 Uppsala, Sweden
    \label{UPPSALA}}
\titlefoot{IFIC, Valencia-CSIC, and D.F.A.M.N., U. de Valencia,
     Avda. Dr. Moliner 50, ES-46100 Burjassot (Valencia), Spain
    \label{VALENCIA}}
\titlefoot{Institut f\"ur Hochenergiephysik, \"Osterr. Akad.
     d. Wissensch., Nikolsdorfergasse 18, AT-1050 Vienna, Austria
    \label{VIENNA}}
\titlefoot{Inst. Nuclear Studies and University of Warsaw, Ul.
     Hoza 69, PL-00681 Warsaw, Poland
    \label{WARSZAWA}}
\titlefoot{Now at University of Warwick, Coventry CV4 7AL, UK
    \label{WARWICK}}
\titlefoot{Fachbereich Physik, University of Wuppertal, Postfach
     100 127, DE-42097 Wuppertal, Germany \\
\noindent
{$^\dagger$~deceased}
    \label{WUPPERTAL}}
\addtolength{\textheight}{-10mm}
\addtolength{\footskip}{5mm}
\clearpage

\headsep 30.0pt
\end{titlepage}

%
\pagenumbering{arabic}                              
\setcounter{footnote}{0}                            %
\large
\section{Introduction}
Mass corrections to the $Z\ra\bb$ coupling are of order $(\mb^2/\MZ^2)$, 
which is too small to be measured at \lep~and \slc. 
For some inclusive observables,
like jet-rates, the effect is enhanced as $(\mb^2/\MZ^2)/\ycut$, where $\ycut$
is the jet resolution parameter~\cite{massycut}. The effect of the \qb-quark 
mass in the production of three-jet event topologies at the $Z$ peak has for
instance already been measured at \lep~and
\slc~\cite{delmbmz,alephmb,opalmb,sldmb}.   
Multi-jet topologies with $b$-quarks appear both as signal and background in
searches and precision measurements at current and future colliders. Their
study, together with that of the gluon emission from massive quarks, is an 
effective tool to probe the fundamental short-distance QCD features of
the Standard Model and is important to test the modelling of $b$ and
light-quark jets available in calculations and generators.

This study generalizes the methods described in references~\cite{delmbmz,MJ} 
and presents the measurement of the 
normalized $n$-jet production partial widths for $Z$-decays into \qb-quark or
light quark pairs:
\beq
      R_{n=2,3,4}^{q}(\ycut) = \left[\frac{\Gamma_n(\ycut)}{\Gamma_{tot}}\right]^{Z\ra\qq}
                   \quad q=b,\ell \; (\ell = uds), 
\label{eq:rnqobs}
\eeq \noindent
depending on the \ycut~value of the \cambridge~jet-clustering 
algorithm~\cite{cambridge} which is used here. 

The effect of the heavy \qb-quark mass on jet rates is studied by 
measuring the double-ratio observable:
\beq
      R_{n=2,3,4}^{b\ell}=R_n^b/R_n^\ell \; .
\label{eq:rnblobs}
\eeq \noindent

The \delphi~data collected during the years 1994 and 1995 at a centre-of-mass 
energy of $\sqrt{s}\approx\MZ$ have been analysed. 
Experimental results are compared to the hadronic final state simulated by the
fragmentation models of \Pythiav~\cite{pythia}, \Herwigv~\cite{herwig}
and \Ariadnev~\cite{ariadne}~and to matrix element (ME) calculations folded 
with a hadronisation correction.
Therefore, the data are corrected for detector and kinematical effects, 
while ME calculations, computed at parton level, 
are corrected for hadronisation.

In order to extract the $b$-quark mass information from \Rnbl~measurements,
massive ME calculations performed in terms of both the pole  
mass $M_b$ and the running mass $m_b(\mu)$ are used. Jet-rate calculations 
are only available to $\Order(\alphas^{2})$~\cite{nlo1,nlo2,nlo3}, therefore
massive  four-jet observables can only be described to leading-order (LO)
accuracy. The $b$-quark mass obtained from \Rfbl~using such LO 
calculations is compared to the three-jet results~\cite{MJ} and to
mass values at threshold~\cite{pdg}  evolved to the $M_Z$ scale using 
Renormalisation Group Equations (RGE). An approximate massless
NLO correction is also tried as an improvement.

The precision of \qb-mass measurements from three-jet events is limited by
systematic uncertainties (hadronisation, \qb-tagging and theory). 
The four-jet observable $R_4^{b\ell}$ has a larger statistical error
but its sensitivity to the $b$-quark mass is higher because, most probably, 
the emission of two gluons is involved. 
The four-jet topology thus provides a complementary measurement in which the 
systematic uncertainties can be 
expected to be partly different. 
In this analysis, flavour jet-rates are measured
using a double-tag technique which measures signal and background
efficiencies from data in a self-calibrating way, reducing the systematics
and allowing for a useful cross-check of previous measurements~\cite{MJ}.




\section{The DELPHI detector}
\delphi~was a hermetic detector located at the \lep~accelerator, with a
superconducting solenoid providing a uniform magnetic field of $1.23$~T
parallel to the beam axis throughout the central tracking device volume.  
A detailed description of its design and performance 
is presented in~\cite{delphi1,delphi2}. 

In the \delphi~coordinate system, the $z$ axis is oriented along the direction
of the electron beam.
The polar angle $\theta$ is measured with respect to the $z$ axis, 
$\phi$ is the azimuthal angle in the plane transverse to the $z$ axis
and $R=\sqrt{x^2+y^2}$ is the radial coordinate. 

The main tracking devices in \delphi~were the silicon Vertex Detector (VD),
a jet chamber Inner Detector (ID) and a Time Projection Chamber (TPC). They
were located in the immediate vicinity of the interaction region to reduce the amount 
of material between the beam and the detector. At a larger distance,
the tracking was completed by a drift chamber Outer Detector (OD) covering
the barrel region ($40^\circ \leq \theta \leq 140^\circ$) and two sets of
drift chambers, FCA and FCB, located in the endcaps.

The VD was the detector closest to the interaction point.
In 1994 and 1995 it consisted of three coaxial cylinders, the inner and outermost
ones consisting of double-sided detectors with orthogonal strips, allowing the measurement
of both $R\phi$ and $z$ coordinates.

Electron and photon identification was provided by electromagnetic calorimeters: 
the High Density Projection Chamber (HPC) in the barrel and a lead-glass
calorimeter (FEMC) in the endcaps.
Hadronic energy was measured in the hadronic calorimeter (HCAL). 


\section{Data analysis}

First, the sample of $Z$ hadronic decays, i.e. $Z\ra\qq$ events was
selected. Then the different jet-topologies were identified using the
\cambridge~jet-clustering algorithm~\cite{cambridge}\footnote{In our analysis,
the values of the ordering and resolution parameters were taken to be equal, $v_{ij}=y_{ij}$.},
and $b$ and light-quark samples were
separated using the \delphi~flavour tagging methods, based on properties of
the long-lived heavy $B$-hadrons. Experimental results were then corrected
for detector and acceptance effects in two different ways, depending on the
observable and topology, as explained in Section~\ref{sec:expcorrection}. 
Matrix element and
event generator predictions were corrected for hadronisation effects from the
parton to the hadron level. 
The parton level is defined as the final state of the parton
shower (in \Pythia~and \Herwig) or dipole cascade (in \Ariadne) in the
simulation, before hadronisation. 
These corrections are discussed in Section~\ref{sec:results}.

\subsection{Event selection}
Total numbers of $1\,484\,000$ and $750\,000$ hadronic $Z$ boson decays,
collected at the $Z$ resonance by \delphi~during the years 1994 and 1995,
respectively, have been 
analysed in order to study mass effects in multi-jet 
topologies\footnote{Earlier 
data samples were not considered as the VD setup was less complete and
resulted in a less precise flavour identification, 
which is crucial for this analysis.}.

Hadronic events were selected in the same way as in reference~\cite{delmbmz}
(see Table \ref{tab:event_selection}, left):
\bi
\item[$\bullet$] Charged and neutral particles were reconstructed as tracks and 
energy depositions in the detector. 
A first selection was applied to ensure a reliable 
determination of their momenta and energies;
\item[$\bullet$] The information from the accepted tracks was combined 
event-by-event and hadronic events 
were selected according to global event properties. 
\ei
Finally, a total sample of $1\,150\,000$ $Z$ hadronic decays was selected.
Then jets were reconstructed with the \cambridge~algorithm. 
In order to reduce the impact of particle losses and wrong 
energy-momentum assignment to jets, 
further kinematical selections were applied,
which were slightly different for each jet topology (see Table
\ref{tab:event_selection}, right).

Simulated events were produced with the \delphi~simulation program \delsim~\cite{delphi2},
based on {\sc Pythia 7.3} tuned to \delphi~data \cite{delphituning}, and were then passed through the same
reconstruction and analysis chain as the experimental data.
The simulated events were reweighted in order to reproduce the measured rates 
of \bb~and \cc-quark pairs arising 
from the gluon splitting processes~\cite{gsplit}
($\gbb=0.00254\pm 0.00051$, $\gcc=0.0296\pm 0.0038$), 
which are significantly larger than those in the standard simulation.
\begin{table}[htb]
\begin{center}
\vspace{7mm}
\begin{tabular}{cc}
\begin{tabular}{|l|l|}
\hline
             & $ p_{ch} \geq 0.1$ GeV/c                         \\
 Charged     & $ 25^{\circ} \leq \theta \leq 155^{\circ}$  \\
 Particle    & $ L \geq 50$ cm                             \\
 Selection   & $ d \leq 5$ cm in $R\phi$ plane             \\
             & $ d \leq 10$ cm in $z$ direction            \\
\hline
\hline
 Neutral     & $E_{cl}^{\mathrm{HPC}}\geq 0.5$ GeV,$40^{\circ} \leq \theta \leq 140^{\circ}$   \\
 Cluster     & $E_{cl}^{\mathrm{FEMC}}\geq 0.5$ GeV,$8^{\circ}\leq\theta\leq 36^{\circ}$\\
 Selection   & $E_{cl}^{\mathrm{FEMC}}\geq 0.5$ GeV,$144^{\circ}\leq\theta\leq 172^{\circ}$\\
             & $E_{cl}^{\mathrm{HAC}} \geq 1$ GeV,   $10^{\circ} \leq \theta \leq 170^{\circ}$ \\
\hline
\hline
             & $N_{ch} \geq 5$                            \\
 Event       & $E_{ch} \geq 15$ GeV                       \\
 Selection   & $|\sum_i q_i|\leq 6, i=1,...,N_{ch}$       \\
             & No particle with $p_{ch} \geq 40$ GeV/c \\
\hline
\end{tabular} &
\begin{tabular}{|c|c|}
\hline
 2-jet                       & $ 45^{\circ}\leq\theta_{thrust}\leq 135^{\circ}$ \\
\hline
 3-jet                       & $ 45^{\circ}\leq\theta_{thrust}\leq 135^{\circ}$ \\
                             & $N_{j}^{ch}\geq 1$                      \\
                             & $E_j\geq 1$ GeV                       \\
                             & $25^{\circ}\leq\theta_{j}\leq 155^{\circ}$\\
                             & $\sum_{ij}\phi_{ij}\geq 359^{\circ}$, $i<j$\\
\hline
 4-jet                         & $ 32^{\circ}\leq\theta_{thrust}\leq 148^{\circ}$ \\
                               & $N_{j}^{ch}\geq 1$                   \\
                               & $E_j\geq 1$ GeV,                     \\
                               & $25^{\circ}\leq\theta_{j}\leq 155^{\circ}$\\
\hline
\end{tabular} \\
\end{tabular}
\end{center}
 \caption{(Left) Particle and event selections: 
  $p_{ch}$ is the momentum of charged particles,
  $L$ their measured track length, 
  $d$ their impact parameter with respect to the interaction point 
  and $q_i$ their charge, 
  $E_{cl}$ is the energy of neutral clusters in the calorimeters, 
  $N_{ch}$ is the number of charged particles 
  and $E_{ch}$ their total energy in the event.
  (Right) Kinematical selections for jets in accepted events: $\theta_{thrust}$ is the polar angle of the thrust of the
  event, $N_j^{ch}$ the charged multiplicity in the jet, $E_j$ the jet energy
  and $\theta_j$ the angle between the jet and the beam axis.
  For three-jet events, an additional planarity cut is applied on the sum 
  of all jet pair angles, $\phi_{ij}$. 
}
\label{tab:event_selection}
\end{table}

\subsection{$b$-tagging}
\label{sec:fourjettag} 
The identification of $b$-quark events in \delphi~was based on the properties
of a $B$-hadron
such as its large mass and the large impact parameter of its decay products.
A jet estimator variable $X_{jet}$ was built as an optimal combination of five
discriminating variables~\cite{newbtag}. 
The most discriminant one was the probability of having all charged particles 
in the jet produced at the event interaction point.
The use of this variable alone 
defined the impact-parameter technique.
The additional variables were used only when a secondary vertex (SV) was
reconstructed. These variables were, for all particles attached to the SV: the
invariant mass, the fraction of the charged jet energy,  
the sum of all transverse momenta and the rapidity of each particle. 
The information from all five variables was combined
into a single estimator $X_{jet}$ in an almost optimal way which provided 
discrimination between heavy and light jets with 
high purity and efficiency. To obtain  $b$(light)-quark enriched samples, 
jets with an estimator value above (below) a given threshold
$X_{jet}\geq X_{jet}^b$ ($X_{jet} < X_{jet}^\ell$) were selected.
To tag events, the value of the two highest $b$-tagging jet variables were
combined into an event estimator,
$X_{ev}=X_{jet}^1+X_{jet}^2$. 

In the present analysis, this approach has been associated to a double-tag
technique~\cite{delphiRb}, 
which measures flavour-tagging efficiencies directly from data. 

Using the two jets with highest $b$-tagging variables as the {\em flavour jets} (jets which are
expected to contain a primary quark) makes no distinction between primary quarks originating
in the $Z$ decay and secondary production of $b$ and $c$-quarks from gluons
($g\ra\bb,\cc$), a process referred to as gluon splitting 
and which constitutes a significant part of the systematic uncertainty in multi-jet
flavour-observables (see Section 4.3).

To reduce the sample contamination from gluon splitting in four-jet events, 
the flavour jets were defined as follows: the most energetic jet in each
event 
is identified as the first flavour jet. Remaining jets are ordered
by angular proximity to it. The closest jet is 
discarded making the hypothesis that it is a gluon coming from the same primary 
quark.
The second $b$(light)-flavour jet is that with the highest \qb-tag (lowest
$b$-tag) estimator among the two remaining jets. 
In this way, energy and angle information is combined to define 
the flavour-jets. As an additional selection, an event is not
classified as \bb~if the most \qb-tagged jet is not among the two most energetic jets;
this last selection reduces the uncertainty from \gbb~and \gcc~by a factor two.
           The effect of the remaining contamination due to
           gluon splitting is included in the gluon-splitting
           uncertainty and is well below the statistical
           uncertainties (Tables~\ref{tab:rnqsystbreak} 
           and~\ref{tab:systbreak}).

\subsection{Overview of the correction method}
\label{sec:expcorrection}
\subsubsection{Event-tag}

To correct the two-jet observable $R_2^{b\ell}$ for detector 
effects and the flavour tagging procedure, 
the event-tag method described in reference~\cite{delmbmz} was used:
\beq
   \Rtwobl = \frac{ 
     [c_B^\ell d_{2B}^\ell + R_{2}^{c\ell} c_B^c d_{2B}^c] -
     [c_L^{\ell} d_{2L}^{\ell} + R_{2}^{c\ell} c_L^c d_{2L}^c] R_2^{b\ell-det} }
   {  c_L^b d_{2L}^b R_2^{b\ell-det} - c_B^b d_{2B}^b },
\label{eq:3jmethod}
\eeq \noindent
where the measured rate $R_2^{b\ell-det}$ is corrected by using purities of the
inclusive samples, $c_Q^q=N_Q^q/N_Q$ 
(the fraction of $\qq$ events tagged in the $Q$ category), 
and detector corrections taken from the
\delsim~simulation, $d_{2Q}^q=R_{2Q}^q/R_2^q$ (where $R_{2Q}^q$ is the
two-jet rate of $\qq$ events tagged as $Q$). 
The factor $R_2^{c\ell} = R_2^c / R_2^{\ell}$ is taken from the simulation. 
Table~\ref{tab:wp2purities} summarizes the number of events selected in the
1994 data in each flavour sample for the chosen working points of 
purity $P_B=c_B^b=98\%$ ($P_L=c_L^{\ell}=73\%$, $L=uds$) 
and efficiency of $\epsilon_B^b = 38\%$ ($\epsilon_L^l=58\%$) 
for \qb-flavoured (light-flavoured) events
(where $\epsilon_Q^q=N_Q^q/N_q$, the ratio of tagged events of a given
flavour to the total number of events of the same flavour),
respectively.

The event-tag method has the advantage of applying the flavour-tagging
procedure only in the inclusive sample, before events are classified into
jet topologies. 

\subsubsection{Double-jet tag}
The event-tag method, if the jet sample is topologically very different from 
the inclusive one, can introduce important biases. To prevent this, in the
$R_4^{b\ell}$ measurement $b$-tagging is applied to jets. 
The observable in Eq.~\ref{eq:rnblobs} is rewritten as: 
\beq
 \label{eq:4jformula}
 \Rfbl = \left( \frac{\Gamma(\Ztoll)}{\Gamma(\Ztobb)} \right) 
         \frac{N_4^b/N_4}{N_4^\ell/N_4} = 
         \left( \frac{1-\Rb-\Rc}{\Rb} \right) \frac{N_4^b}{N_4^\ell}.
\eeq 
\noindent
The global normalisation can be obtained directly from the world average values of \Rb~and
\Rc~\cite{pdg}:
\beqn
\label{eq:rbrcvalues}
    \Rb &=& 0.21629 \pm 0.00066, \\ 
    \Rc &=& 0.1721  \pm 0.0030,  \nonumber
\eeqn 
\noindent
which implies a $\pm 6\tcperthousand$ uncertainty on \Rfbl.
A double-tag technique is used: the total number of four-jet events, $N_4$, 
the corresponding numbers for a given flavour $N_4^q$, $q=b,udsc$, 
and the tagging efficiencies $\epsilon_B^b$ and $\epsilon_{UDSC}^{udsc}$ 
are obtained from comparing the number of four-jet events where  
two jets are tagged as $b$ or $udsc$ to the number of events where a
single jet is tagged. This is done by solving the following set of equations:
\beqn
\label{eq:dt1}
  {\mathcal{N}}_4&=&\left\{N_4^b\epsilon_h^b+(N_4-N_4^b)\epsilon_h^{non-b}\right\},\\
\label{eq:dt2}
  \frac{1}{2}{\mathcal{N}}_{4B}&=&\left\{N_4^b\epsilon_h^b\epsilon_B^b
  +(N_4-N_4^b)\epsilon_h^{non-b} \epsilon_B^{non-b} \right\}, \\ 
\label{eq:dt3}
  {\mathcal{N}}_{4BB}&=&\left\{N_4^b\epsilon_h^b\epsilon_{BB}^b
  +(N_4-N_4^b)\epsilon_h^{non-b}\epsilon_{BB}^{non-b}\right\},  
\eeqn 
\noindent
and equivalent equations for
the $udsc$-tagged samples. 
The left hand side of these equations are the measured
quantities. 
${\mathcal{N}}_4$ is the number of measured four-jet events.
For each event the two jets which are most likely to contain a primary
quark (flavour jets, see above) 
are selected and the flavour identification is
done independently for both jets: ${\mathcal{N}}_{4B}$ is the number of jets
tagged as $B$ (with a maximum possible value of $2{\mathcal{N}}_4$, two from
each event) and ${\mathcal{N}}_{4BB}$ is the number of events where the two 
flavour jets are simultaneously tagged as $B$.  
With this method, the jet-rates $R_4^b$ and $R_4^\ell$ are measured
independently, together with the efficiencies $\epsilon_B^b$ and
$\epsilon_{UDSC}^{udsc}$. 
To accomplish this, double-jet tagging efficiencies $\epsilon_{QQ}^q$ are
related to the single jet-tagging efficiencies through correlation factors
defined from $\epsilon_{QQ}^q=N_{QQ}^q/N^q \equiv \epsilon_Q^q \epsilon_Q^q (1+\rho_Q^q)$.
Here,
charm-events have been included in the $udsc$-tagged category: 
the light-quark content $N_4^\ell$ is extracted from $N_4^{udsc}$ 
after dividing by a factor
$(1+N_4^c/N_4^\ell)$ obtained from Monte Carlo event generators. Only hadronic
event-selection efficiencies for each flavour, $\epsilon_h^q=N_q^{sel}/N_q$, 
mistagging efficiencies, 
$\epsilon_{Q}^{non-q}$ and $\epsilon_{QQ}^{non-q}$, and flavour
correlations for $b$ and light-tagging are computed from the simulation.

This procedure can be easily generalised to cover $n=2,3$-jet topologies 
in order to measure both jet-rates ($R_n^b,R_n^\ell$) and the 
double-ratios ($R_{2,3}^{b\ell}$) independently. 
Due to the $6\tcperthousand$ uncertainty from the global
normalisation, the double-tag measurements for $R_{2,3}^{b\ell}$ are less
precise than the corresponding event-tag result. However, they serve as a useful
cross-check both of the final result and on the consistency between data and
simulation for the flavour-tagging efficiencies.  

\begin{figure}
\centering
  {\includegraphics[width=0.58\linewidth]
  {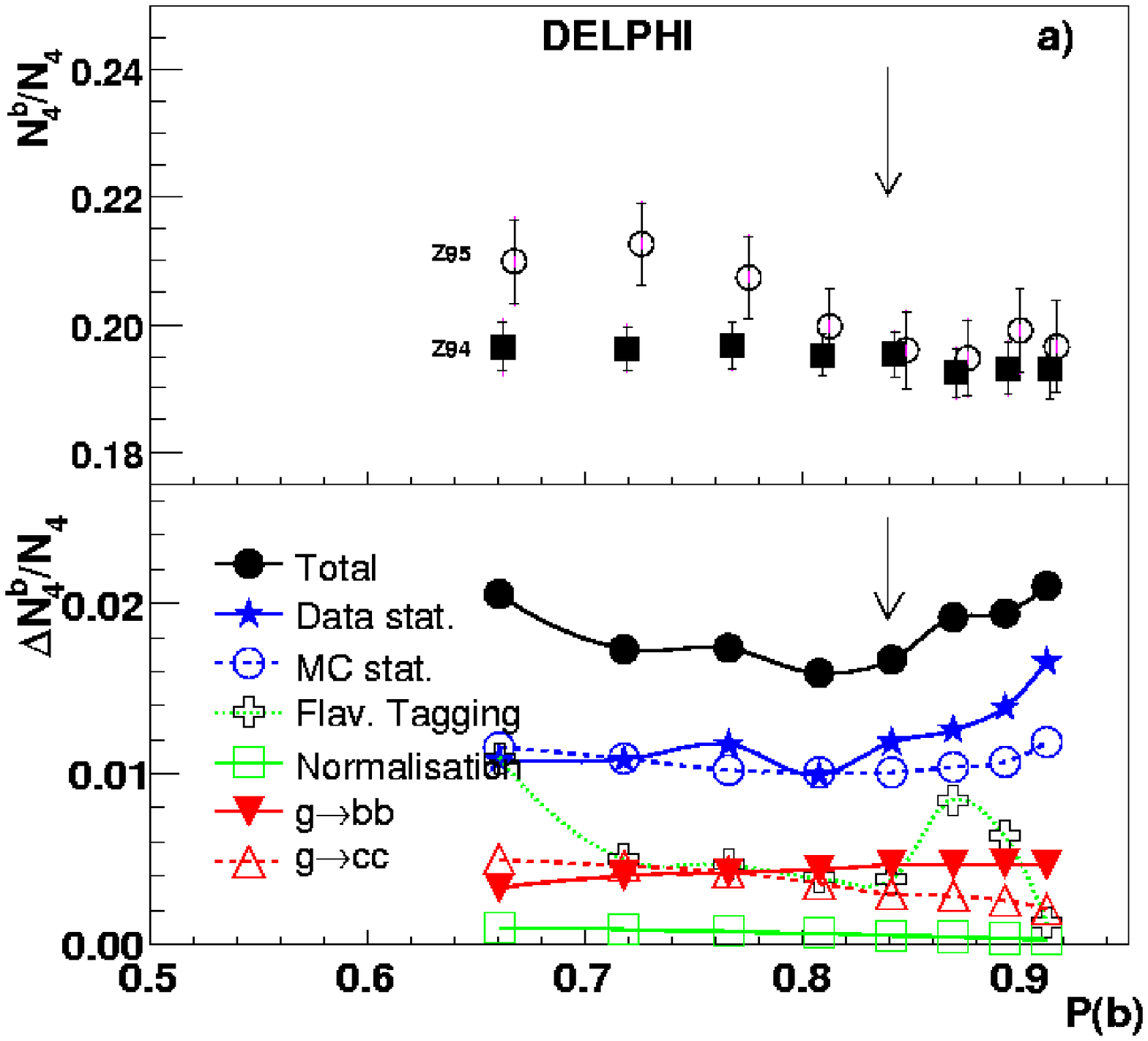}}\\
  \vspace{0.50cm}
  {\includegraphics[width=0.58\linewidth]
  {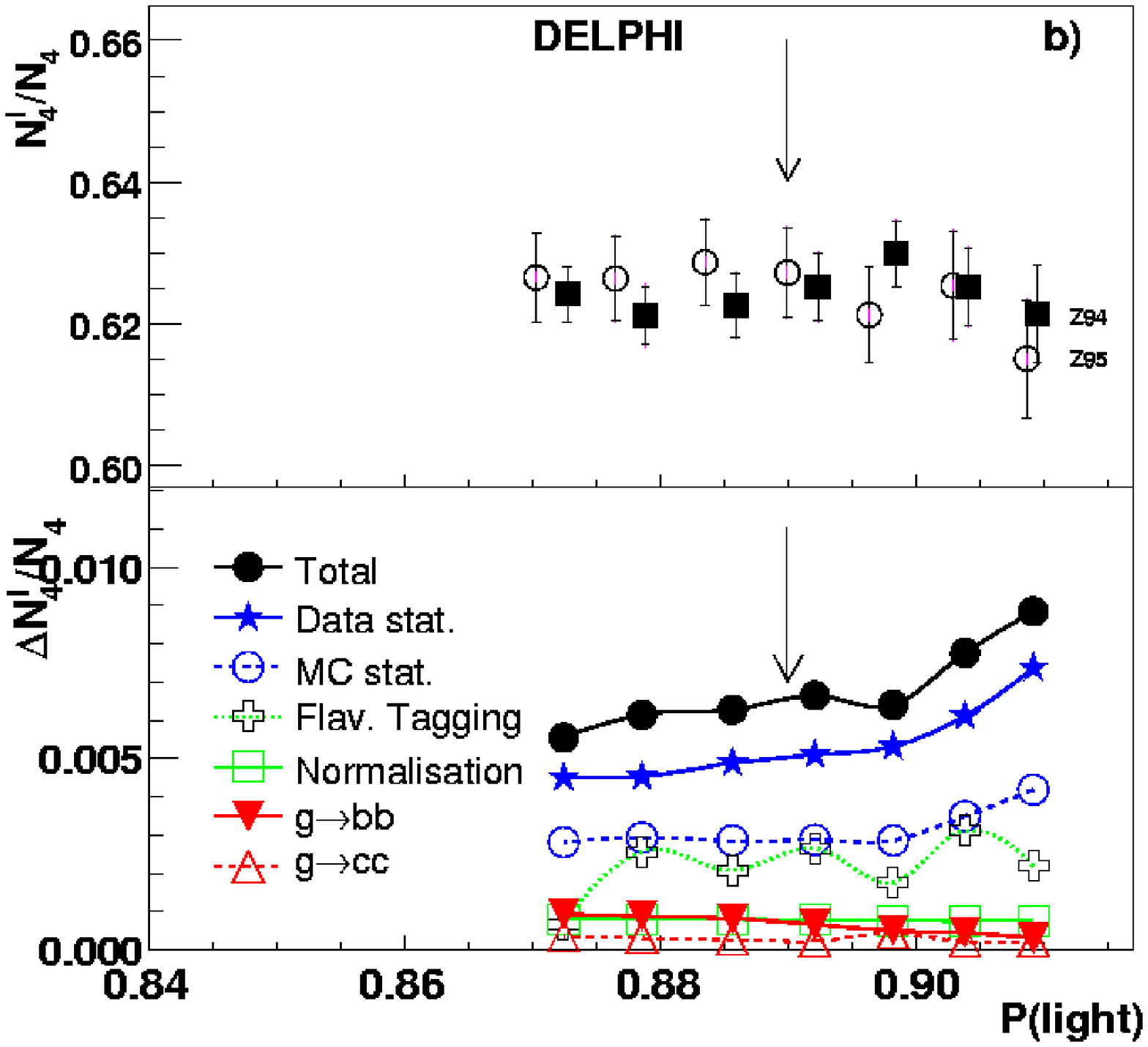}}
\caption{Four-jet rate and its uncertainty as a function of 
        (a) the \qb-purity and (b) the light-purity ($\ell = udsc$, 
        $\ycut=0.0065$). Chosen working points are marked with arrows, and
        correspond to efficiencies of 
	$\epsilon_B^b=35\%$ and $\epsilon_L^{\ell}=86\%$, respectively. 
        The statistical (data and simulation) and total uncertainties are shown.} 
 \label{fig:fourjstab}
\end{figure}

Results with this method
have a better stability with respect to the 
value of the flavour-tagging threshold, 
and are more consistent with each other\footnote{From the relation
$N_4=N_4^b+N_4^{udsc}$, the double-ratio $R_4^{b\ell}$ can be 
obtained independently in two ways, starting either from $R_4^b$ or
$R_4^\ell$.}.
The flavour composition of the 1994 sample is shown as an example in 
Tables~\ref{tab:wp2purities} and~\ref{tab:wppurities}. 
The stability obtained in the case of the four-jet rates is shown in Figure
\ref{fig:fourjstab} for the 1994 and 1995 data samples.
\begin{table}[htb]
\begin{center}
\vspace{7mm}
\begin{tabular}{|c|c|c|c|c|c|}
\hline
\multicolumn{6}{|c|}{ Event-tag method} \\
\hline
Flavour  & Inclusive &    2 jets & cut               & Purity & Efficiency \\
\hline
\hline
 $B$      &    111440 & ~75147 & $X_{ev}\geq 1.10$ & $98\%$ & $38\%$ \\
 $L$      &    678282 & 414912 & $X_{ev}  <  0.40$ & $73\%$ & $58\%$ \\
\hline
\end{tabular}
\vspace{0.5cm}
\end{center}
\caption{Flavour composition of the 1994 sample ($\ycut\!=0.0065$). 
  The number of
  events in the inclusive and 2-jet samples are shown separately for $B$ and
  $L=uds$ tagged events for the chosen flavour-tagging working
  points. Purity and efficiency are also shown. 
  Similar numbers were found with the 1995 data.}
\label{tab:wp2purities}
\end{table}
\begin{table}[!htb]
\begin{center}
\vspace{7mm}
\begin{tabular}{|c|c|c|c|c|c|}
\hline
\multicolumn{6}{|c|}{ double-tag method} \\
\hline
Topology & $Q$ & $QQ$ &   cut       & Purity & Efficiency \\
\hline
\hline
 2 jets ($B$)& 136228 & ~35187  & $X_{2j}\geq +0.33$   & $92\%$ & $57\%$ \\
 2 jets ($L$)& 640757 & 243716 & $X_{2j}  <  -0.92$   & $95\%$ & $78\%$ \\
\hline
 3 jets ($B$)& ~66034 & ~15356 & $X_{3j}\geq +0.19$   & $87\%$ & $53\%$ \\
 3 jets ($L$)& 362396 & 147605 & $X_{3j} <   -0.64$   & $93\%$ & $84\%$ \\
\hline
 4 jets ($B$)&  10720 & ~2191 & $X_{4j}\geq +0.05$   & $84\%$ & $35\%$ \\
 4 jets ($L$)&  91042 &  36773 & $X_{4j} <   -0.64$   & $89\%$ & $86\%$ \\
\hline
\end{tabular}
\vspace{0.5cm}
\end{center}
\caption{Flavour composition of the 1994 sample ($\ycut\!=0.0065$) 
tagged as $n$-jet \qb-quark ($B$) and $udsc$-quark events ($L$) 
for the different jet topologies analysed, $n=2,3$ and $4$ jets. 
Four-jet tagging uses the method described in Section~\ref{sec:fourjettag}
for the definition of flavour jets. Similar numbers were found with the 1995 data.}
\label{tab:wppurities}
\end{table}

\section{Results}
\label{sec:results}

The single-flavour jet rates $R_n^q$, $n=2,3,4$-jets, 
and the four-jet observable $R_4^{b\ell}$, are
measured with the double-tag technique, 
while the two-jet observable $R_2^{b\ell}$ is 
measured using the event-tag method described in~\cite{delmbmz}. 
A description of the experimental uncertainties
considered in the analysis is given in Section~\ref{sec:expsyst}. Theoretical
uncertainties, arising in the comparison between ME predictions and the
four-jet observable, are discussed 
in Sections~\ref{sec:4jettheo} and~\ref{sec:theosyst}.

\subsection{Single jet-rates, $R_n^q$}
\begin{figure}[hbtp]
\centering
  \includegraphics[width=0.495\linewidth]
  {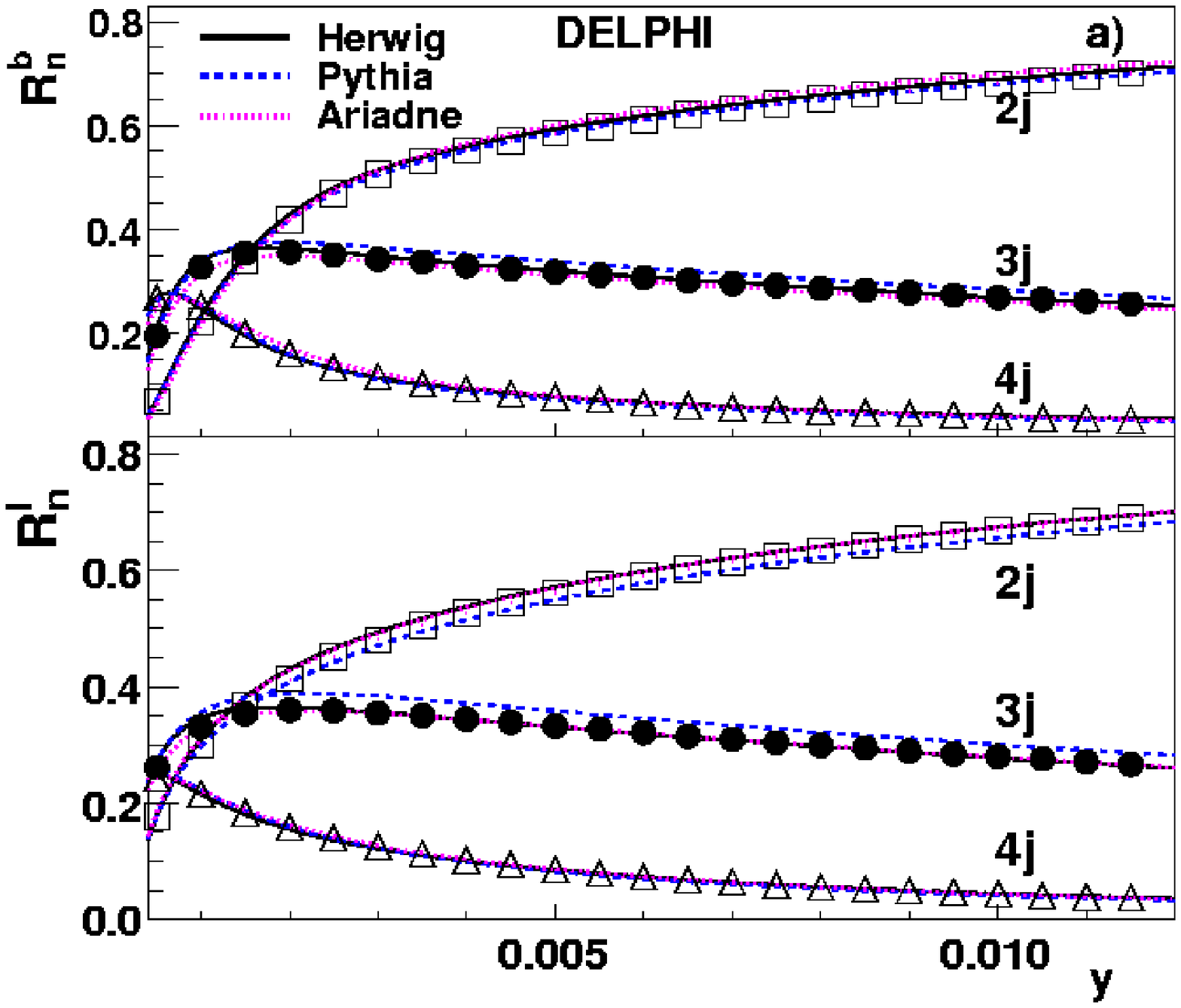}
  \includegraphics[width=0.495\linewidth]
  {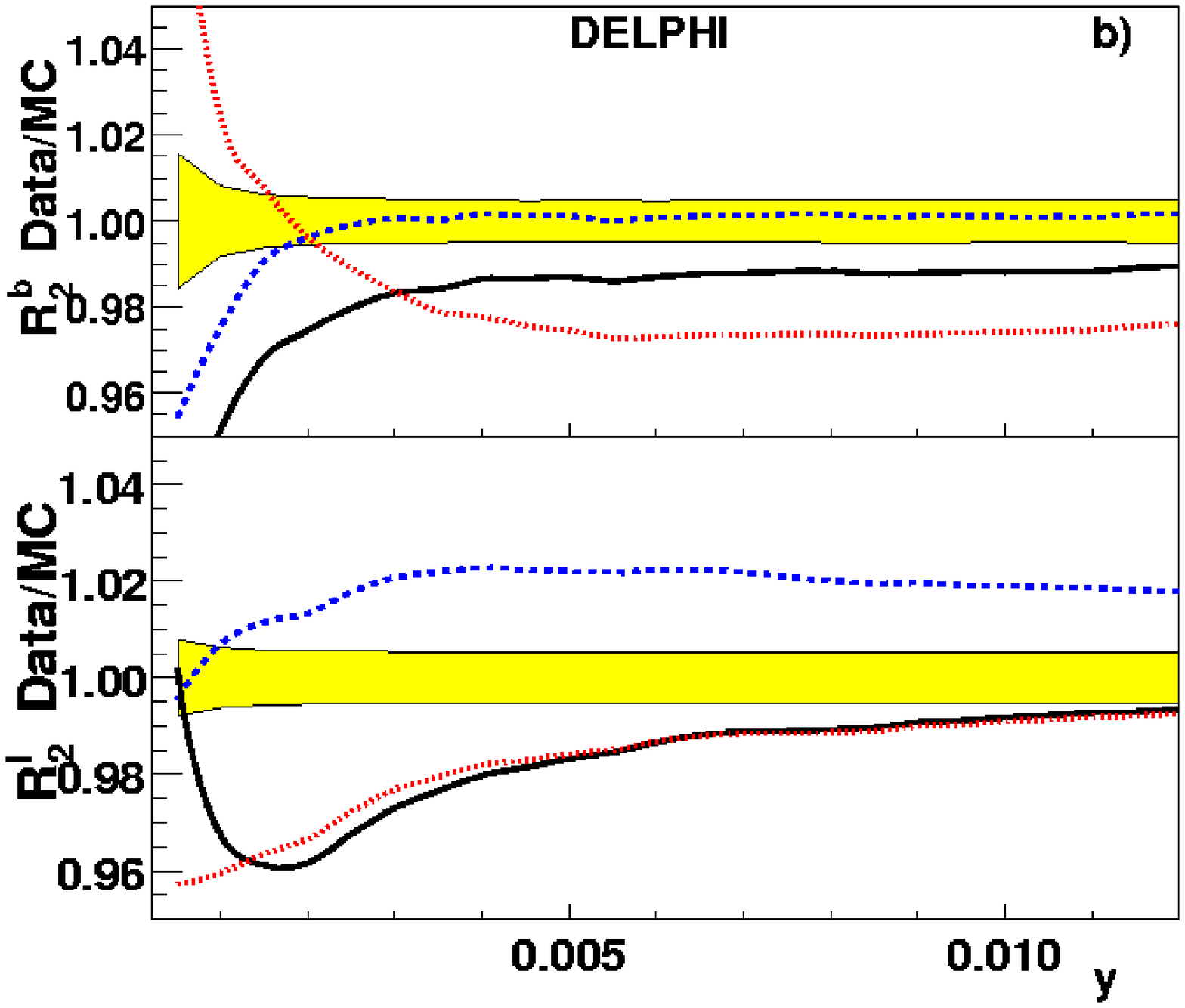}
  \includegraphics[width=0.495\linewidth]
  {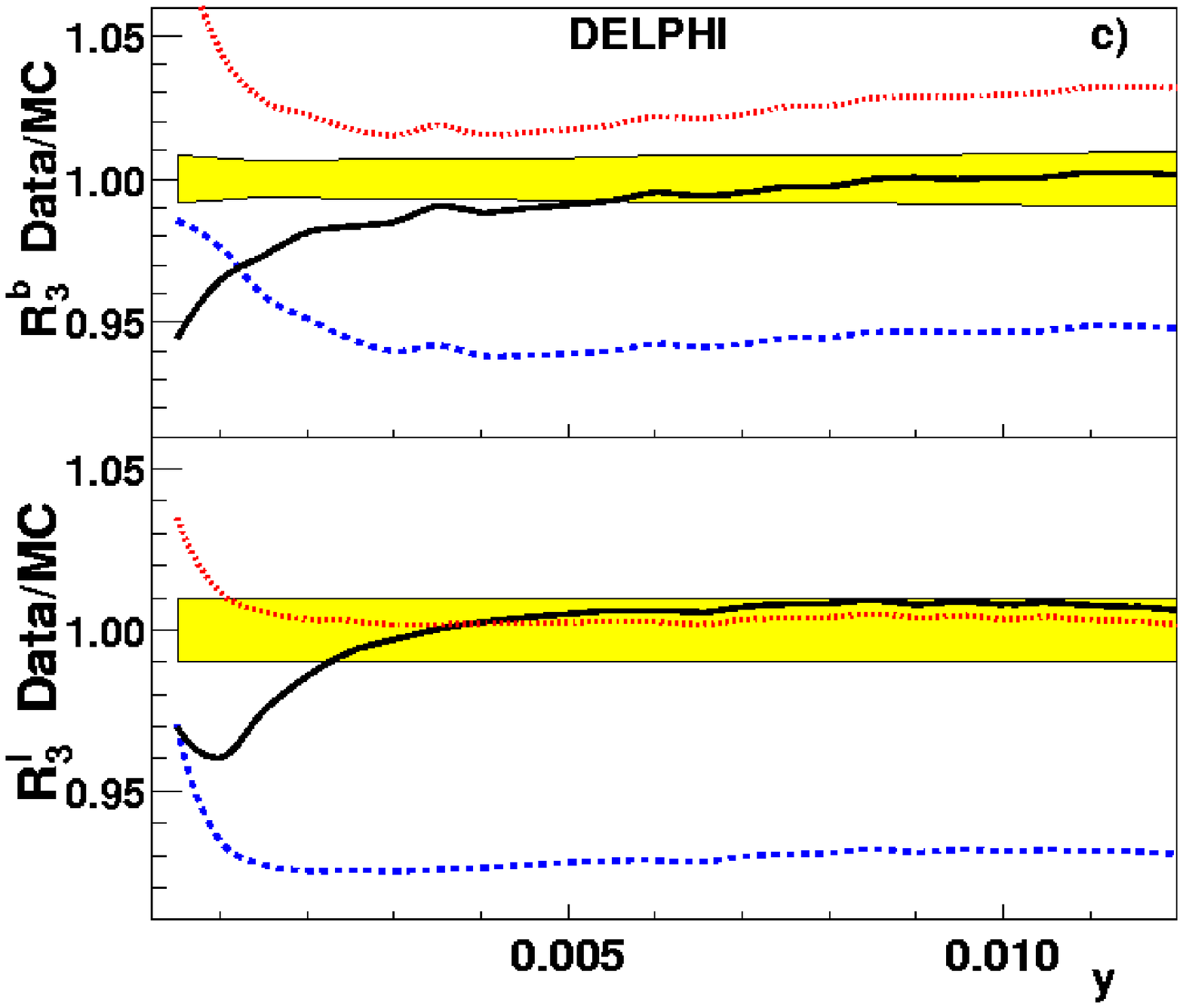}
  \includegraphics[width=0.495\linewidth]
  {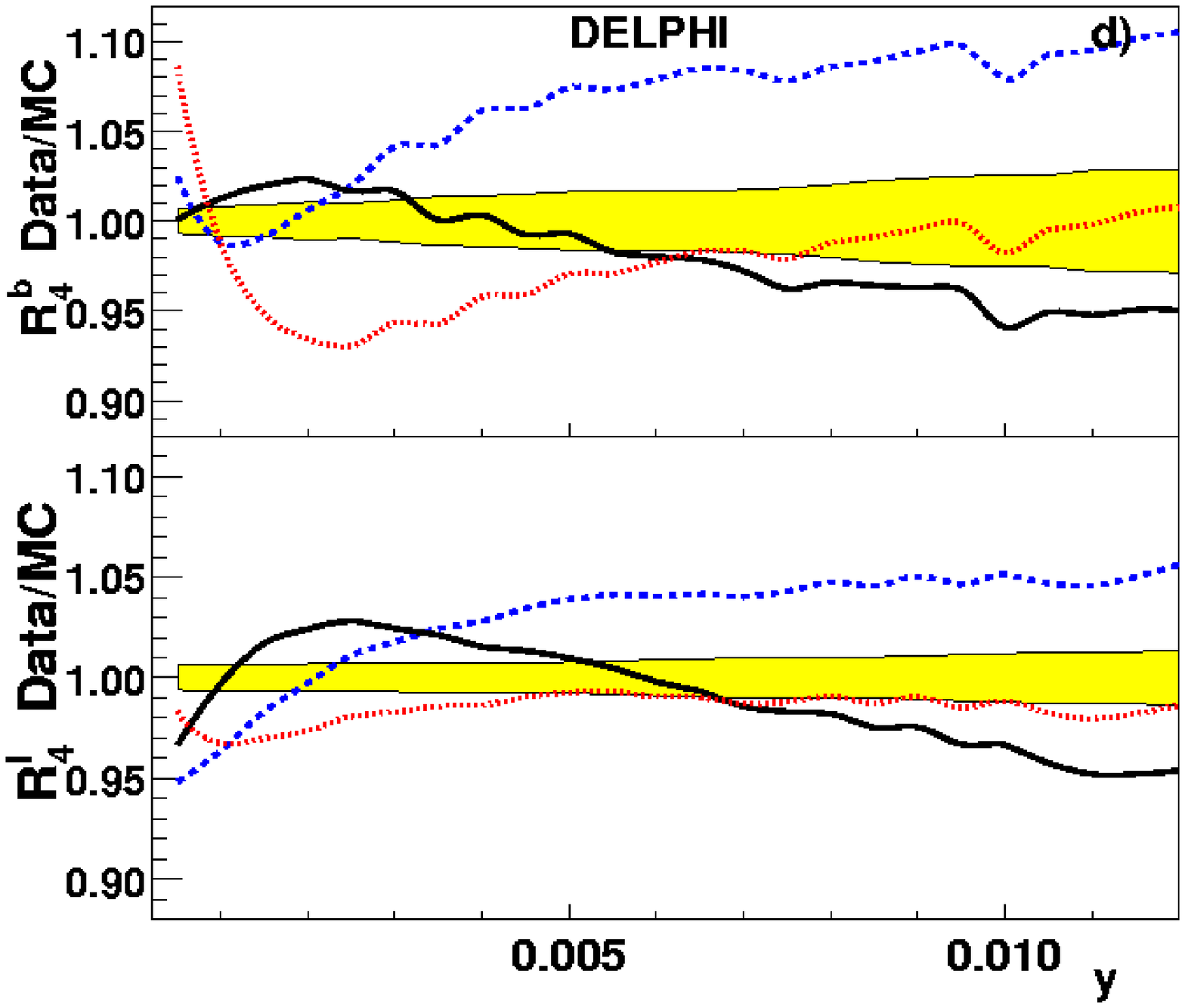}
 \caption{ Comparison between the measured \qb~and $\ell=uds$ jet-rates and
 predictions from the \Pythiav, \Herwigv~and \Ariadnev~generators.
 b-d) Ratio of data to the different generators.
The shaded area shows the one standard deviation relative uncertainty 
(statistical and systematic added in quadrature) of the experimental measurement.
 }  
 \label{fig:jbl_generators}
\end{figure}

The measured $R_n^q$ rates ($n=2,3,4$ jets, $q=b$ or $\ell = uds$) 
are shown in Figure~\ref{fig:jbl_generators}a 
together with predictions from the \Pythiav,
\Herwigv~and \Ariadnev~generators tuned to \delphi~data~\cite{delphituning}
(see Section~\ref{sec:theosyst} for the choice of the \qb-quark mass parameter
in the generators). The detailed breakdown of the uncertainties of the
measured jet-rates is shown in Table~\ref{tab:rnqsystbreak}. 
The $R_3^{\ell}$ measurements in 1994 and 1995 were found to be incompatible
with each other at the two standard deviations level, indicating that some 
systematic effect was not taken into account in the three-jet light-quark rate. 
The systematic tagging uncertainty in $R_3^{\ell}$ was increased in order to
fully cover this difference.
           Only the uncertainty in $R_3^{\ell}$ was increased since the
           $b$-tagging was developed from 2-jet events yielding
           reliable $R_2^{\ell}$ results, and in 4-jet events the
           $b$-tagging applies different cuts on angle and energy.
The consistency of the experimental results and the prediction from the three
event generators is shown in Figures~\ref{fig:jbl_generators}b-c: the
\Herwigv~and \Ariadnev~generators provide a reasonable description of the
six observables in the region of \ycut~between $0.001$ and $0.010$. 
\Pythiav~gives the best description of $R_2^b$, but is inconsistent with the
other jet-measurements at the three standard deviations level.
%
\begin{table}[htb]
\begin{center}
\vspace{7mm}
\begin{tabular}{|c|c|c|c|c|c|c|}
\hline
                        &$R_2^b$      &$R_2^\ell$   & $R_3^b$     & $R_3^\ell$  & $R_4^b$     & $R_4^\ell$\\
\hline
      Value             &$0.6224$     & $0.6034$    &$0.3004$     &  $0.3150$   &$0.0598$     &$0.0676$  \\ 
\hline
Statistical (data)      &$\pm 0.0019$ &$\pm 0.0008$ &$\pm 0.0016$ &$\pm 0.0006$ &$\pm 0.0007$ &$\pm 0.0004$ \\  
Statistical (sim.)      &$\pm 0.0012$ &$\pm 0.0005$ &$\pm 0.0009$ &$\pm 0.0004$ &$\pm 0.0006$ &$\pm 0.0003$ \\   
\hline
       Tagging          &$\pm 0.0004$ &$\pm 0.0008$ &$\pm 0.0006$ &$\pm 0.0025$ &$\pm 0.0001$ &$\pm 0.0002$ \\   
     Normalisation      &$\pm 0.0018$ &$\pm 0.0030$ &$\pm 0.0009$ &$\pm 0.0016$ &$\pm 0.0002$ &$\pm 0.0003$ \\   
        \gbb            &$\pm 0.0003$ &$<0.0001$    &$\pm 0.0009$ &$<0.0001$    &$\pm 0.0003$ &$<0.0001$ \\   
        \gcc            &$\pm 0.0006$ &$\pm 0.0002$ &$\pm 0.0010$ &$<0.0001$    &$\pm 0.0002$ &$<0.0001$ \\   
\hline
    Total systematics   &$\pm 0.0020$ &$\pm 0.0031$ &$\pm 0.0017$ &$\pm 0.0030$ &$\pm 0.0004$ &$\pm 0.0004$ \\    
    Total statistical   &$\pm 0.0023$ &$\pm 0.0009$ &$\pm 0.0018$ &$\pm 0.0007$ &$\pm 0.0009$ &$\pm 0.0005$\\    
\hline
\hline
 Total uncertainty      &$\pm 0.0030$&$\pm 0.0033$&$\pm 0.0025$&$\pm 0.0031$&$\pm 0.0010$&$\pm 0.0006$\\  
\hline
\end{tabular}
\end{center}
\caption{Breakdown of uncertainties for the $R_n^q$ jet-rate measurements
  at a reference $\ycut=0.0065$. The definition of each systematic
  contribution is given in Section~\ref{sec:expsyst}.}
\label{tab:rnqsystbreak}
\end{table}

\subsection{Double-ratios, $R_n^{b\ell}$}
\begin{figure}[!htb]
\centering
 \includegraphics[width=0.70\linewidth]
  {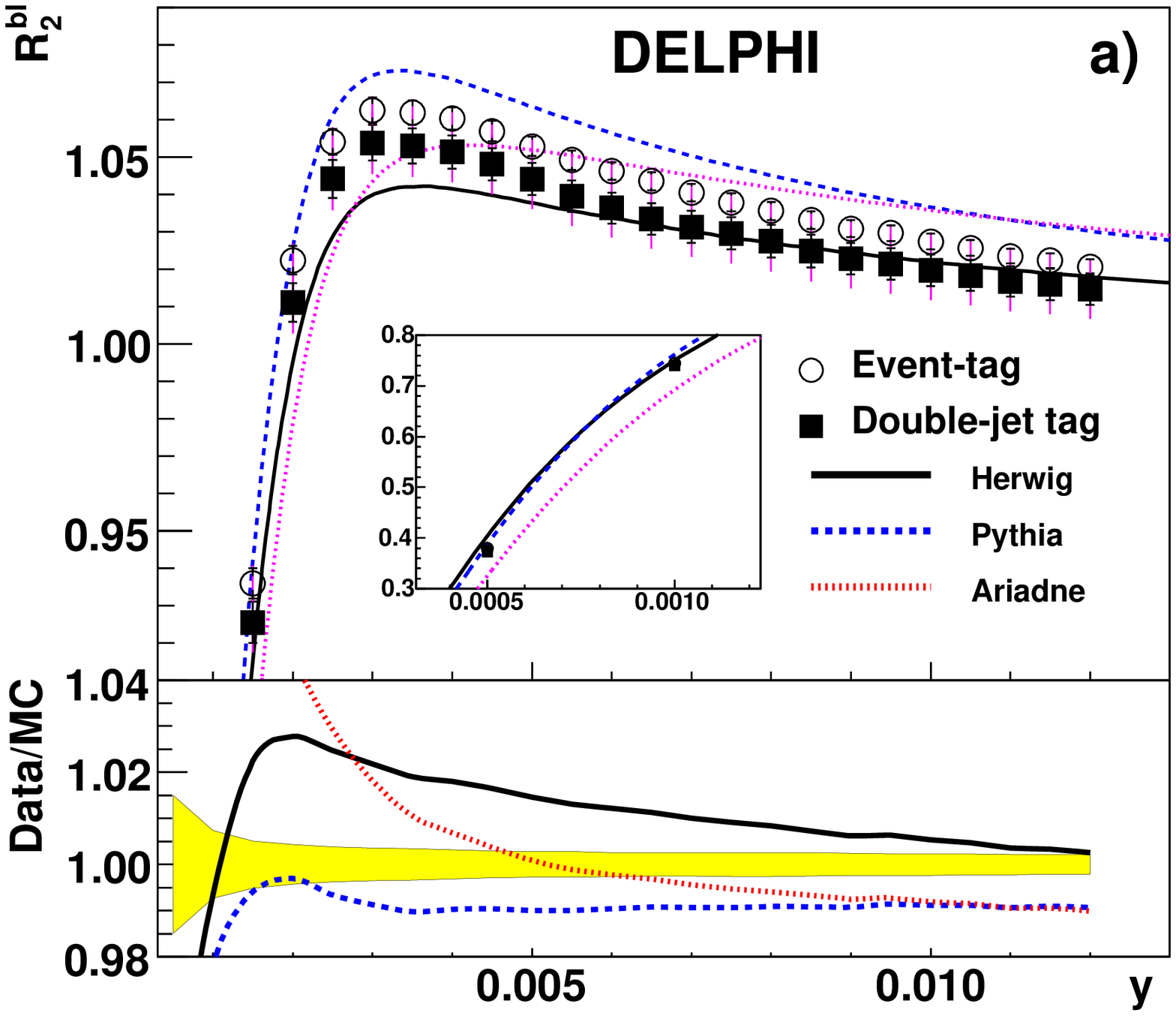}
 \includegraphics[width=0.70\linewidth]
  {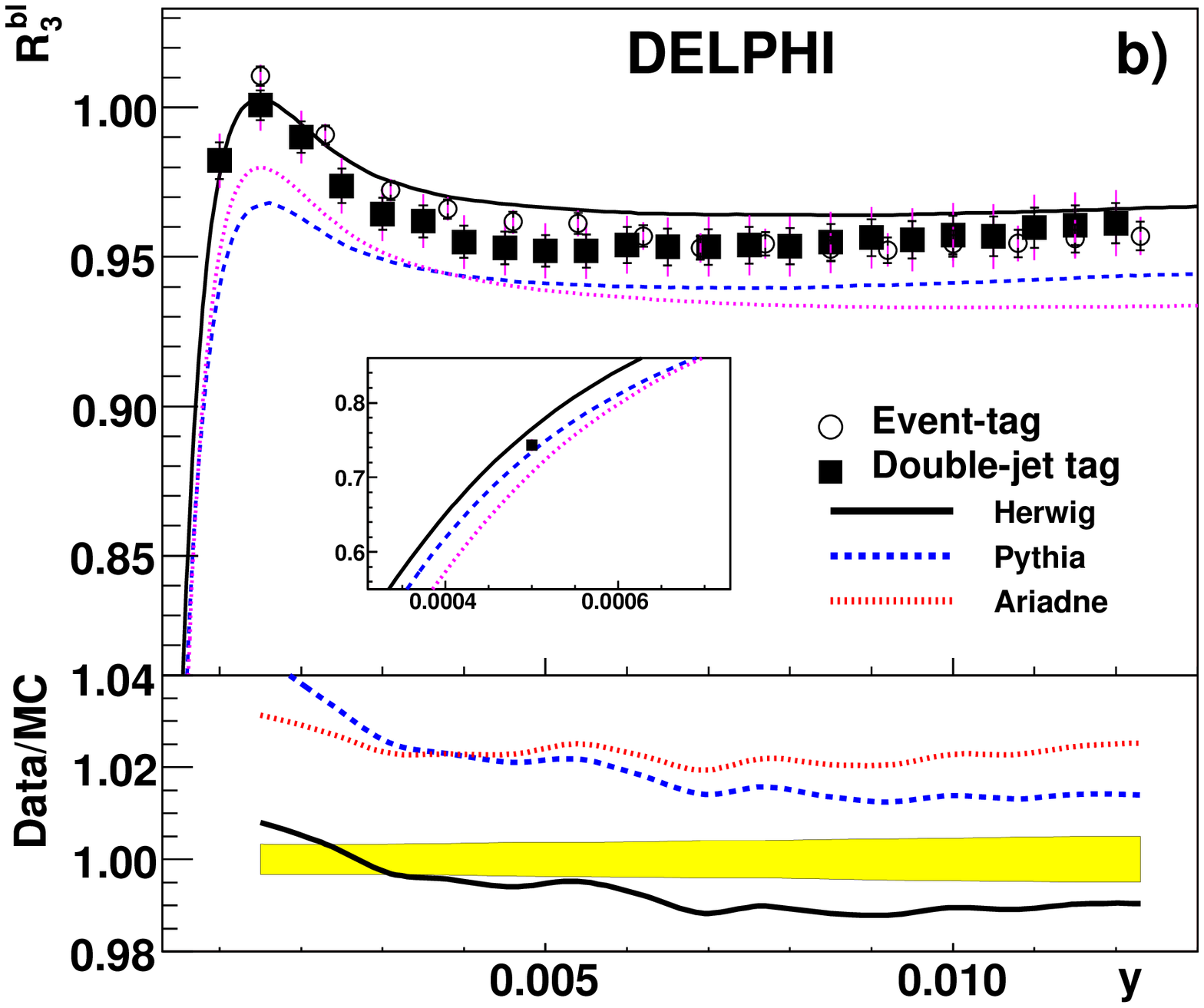}
 \caption{Comparison between the event-tag (empty circles) and double-tag  
          (full squares) techniques for the measured 
          (a) \Rtwobl~and (b) \Rthreebl~observables. 
          The event-tag result of \Rthreebl~is taken from~\cite{MJ}.
          The combined statistical (inner bars) and total uncertainty of the
          experimental data are shown. 
          The results are compared to the predictions from the
          \Herwigv~(solid), \Pythiav~(dashed), and \Ariadnev~(dotted) 
          event generators.
          The lower insets of the plots show the ratio of data to the
          different generators.
          Also shown as the shaded area is the one standard deviation relative
          uncertainty (statistical and systematic added in quadrature) of
          the data. 
}  
 \label{fig:23j_generators}
\end{figure}

\begin{figure}[!htb]
\centering
 \epsfig{file=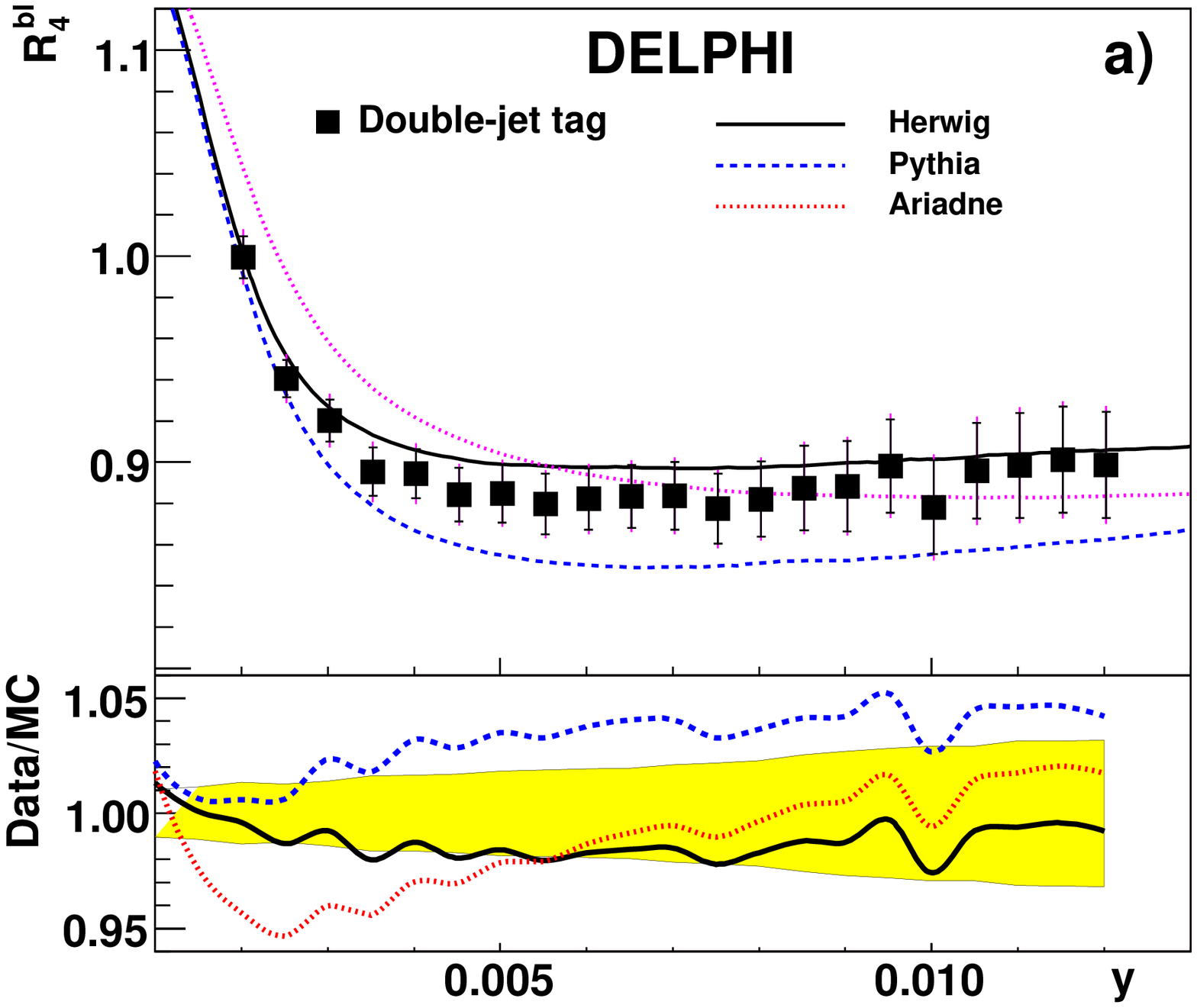, width=0.70\linewidth}
 \epsfig{file=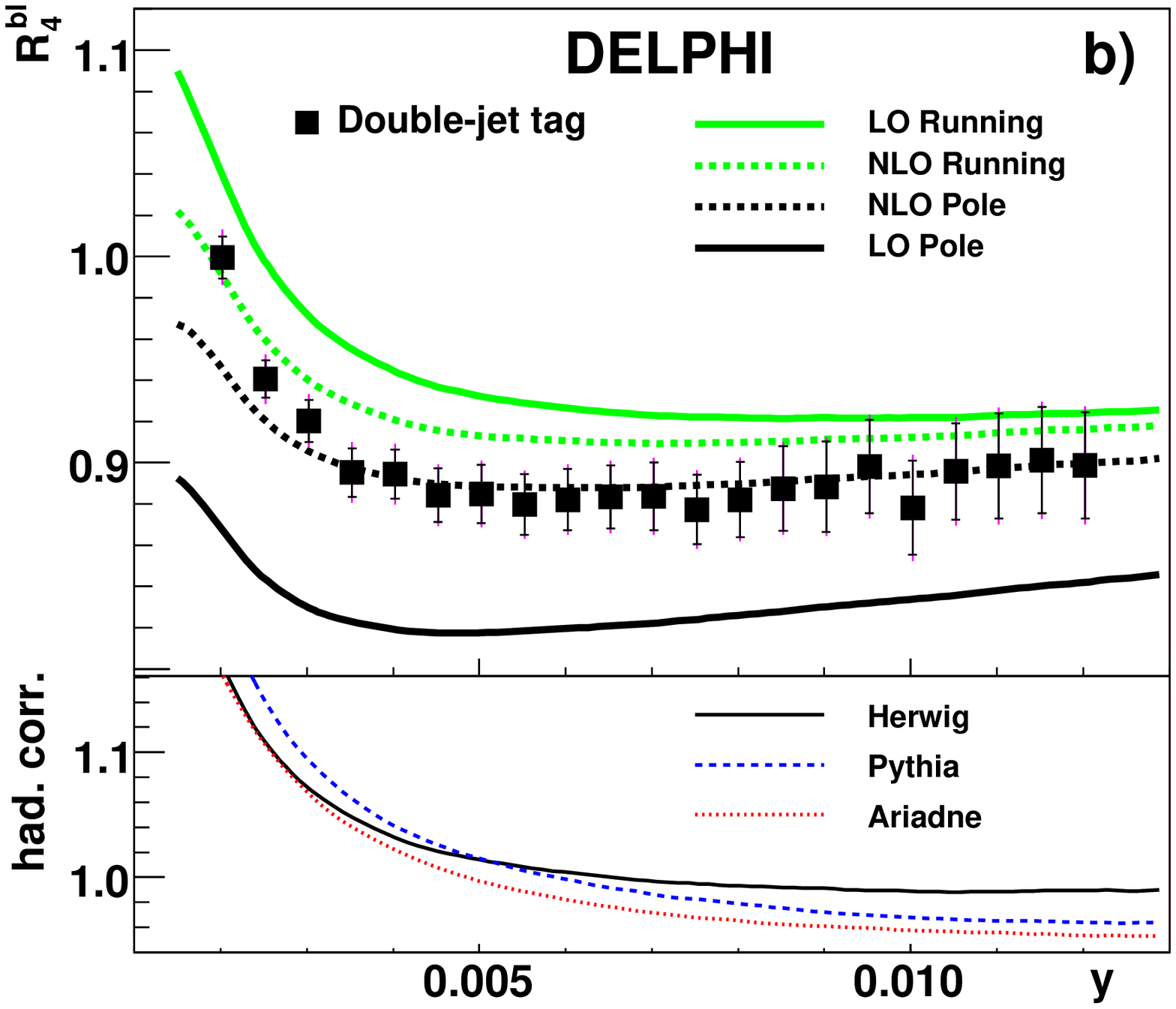, width=0.70\linewidth}
  \caption{ (a) Comparison between the \Rfbl~measured with a double-tag
         technique and predictions from the 
         \Herwigv~(solid), \Pythiav~(dashed), and \Ariadnev~(dotted) 
         event generators. 
          The combined statistical (inner bars) and total uncertainty of the experimental data are shown.
          The lower inset shows the relative deviation of the
          models to the data. Also shown as the shaded area is the total one
          standard deviation relative
          uncertainty (statistical and systematic added in quadrature) of
          the data. Below $\ycut=0.002$, the flavour tagging procedure fails
          and data results from the 1994 and 1995 data samples are not consistent with each
          other. 
         (b) Comparison between the measured
         \Rfbl~and theoretical predictions: massive LO predictions and
         approximate (massless) NLO corrections for the pole and running
         $b$-quark mass definitions. Reference \qb-quark masses were obtained by
         evolving the average of low energy measurements $\mb(\mb)=4.20
         \pm0.07~\GeVcTwo$~\cite{pdg}  
         to the $M_Z$ scale as explained in Section \ref{sec:4jettheo}.
         Hadronisation corrections, used to
         correct ME calculations, are shown for the three generators in the
         lower inset.
} 
 \label{fig:4j_generators}
\end{figure}

The measured double-ratios \Rnbl~($n=2,3,4$ jets) are shown 
in Figures~\ref{fig:23j_generators}-\ref{fig:4j_generators} 
together with predictions from the \Pythiav, \Herwigv~and \Ariadnev~generators 
tuned to \delphi~data~\cite{delphituning} (see Section~\ref{sec:theosyst} for
the choice of the \qb-quark mass parameter in the generators).

Results for \Rtwobl~and \Rthreebl~from the
event-tag and double-tag methods are shown in Figure~\ref{fig:23j_generators}
(event-tag results for \Rthreebl~are taken from~\cite{MJ}). 
\Rtwobl~is not described well by either of the generators in the full 
\ycut~range. 
In all cases, both methods give consistent results 
within one standard deviation. A better experimental precision is found with
the event-tag, because the global normalisation uncertainty is absent in this
case and because flavour-tagging uncertainties 
cancel to first order in the products $c_Q^q d_{nQ}^q$ 
(see Eq.~\ref{eq:3jmethod}). 
Statistical uncertainties in the event-tag result are also smaller, as more data
events are considered and as statistical fluctuations are partially reduced in
the ratios of the jet and inclusive samples. The detailed breakdown of the
uncertainties of the measured double-ratios is shown 
in Table~\ref{tab:systbreak} for the event-tag method.  
\begin{table}[htb]
\begin{center}
\vspace{7mm}
\begin{tabular}{|c|c|c|c||c|c|c|}
\hline
 \multicolumn{4}{|c||}{~} & \multicolumn{1}{|c|}{LO ($\GeVcTwo$)} & \multicolumn{2}{|c|}{NLO ($\GeVcTwo$)}\\ 
\hline
                        & \Rtwobl      & \Rthreebl~\cite{MJ}& \Rfbl   & $\mb(\MZ)=\Mb$ & $\mb(\MZ)$ & $\Mb$ \\
\hline
           Value        & $1.0440$     &   $0.9570$    & $0.883$ & $3.76$ &      $3.46$ &       $5.07$      \\
\hline
Statistical (data)      & $\pm 0.0021$ &   .    &$\pm 0.012$ & $\pm 0.25$ & $\pm 0.27$ &  $\pm 0.35$ \\
Statistical (sim.)      & $\pm 0.0012$ &   .    &$\pm 0.010$ & $\pm 0.20$ & $\pm 0.22$ &  $\pm 0.28$ \\
\hline
Tagging                 & $\pm 0.0009$ &   .    &$\pm 0.003$ & $\pm 0.07$ & $\pm 0.08$ &  $\pm 0.10$ \\
Normalisation           & $\pm 0.0005$ &   -    &$\pm 0.005$ & $\pm 0.11$ & $\pm 0.12$ &  $\pm 0.16$ \\
          \gbb          & $\pm 0.0007$ &   .    &$\pm 0.005$ & $\pm 0.10$ & $\pm 0.10$ &  $\pm 0.13$ \\
          \gcc          & $\pm 0.0003$ &   .    &$\pm 0.003$ & $\pm 0.06$ & $\pm 0.06$ &  $\pm 0.08$ \\
\hline
    Total systematics   & $\pm 0.0013$ &   $\pm 0.0027$    &$\pm 0.008$ & $\pm 0.17$ & $\pm 0.19$ &  $\pm 0.24$ \\
    Total statistical   & $\pm 0.0024$ &   $\pm 0.0037$    &$\pm 0.015$ & $\pm 0.32$ & $\pm 0.35$ &  $\pm 0.46$ \\
\hline
\hline
Total experimental      & $\pm 0.0027$ &   $\pm 0.0046$    &$\pm 0.017$ & $\pm 0.36$ & $\pm 0.40$ &  $\pm 0.52$ \\  
\hline
\hline
 Modelling                &     -        &   -    &   -        & $\pm 0.22$ & $\pm 0.24$ &  $\pm 0.32$\\   
 Theoretical            &     -        &   -    &   -        & $\pm 0.90$ & $\pm 0.44$ &  $\pm 0.57$\\
\hline
\hline
 Total uncertainty      & $\pm 0.0027$ &   $\pm 0.0046$    &$\pm 0.017$ & $\pm 0.99$ & $\pm 0.64$ & $\pm 0.83$\\
\hline
\end{tabular}
\end{center}
\caption{Breakdown of uncertainties for the $R_n^{b\ell}$ $(n=2,3,4)$ 
  double-ratio measurements. The three-jet result is
  taken from~\cite{MJ} and shown here for completeness. The two and three-jet
  measurements are based on the event-tag method, while \Rfbl~uses the
  double-tag technique as explained in Section~\ref{sec:expcorrection}. 
  The $b$-mass values (running and pole) extracted from \Rfbl~at
  reference $\ycut=0.0065$ are also shown, both for the massive LO 
  ($M_b=\mb(\MZ)$ ) and approximate NLO calculations. 
  Experimental and modelling uncertainties (experimental tuning and 
  hadronisation model in the simulation) are detailed separately.}  
\label{tab:systbreak}
\end{table}


The \Rfbl~result with the double-tag method is shown in
Figure~\ref{fig:4j_generators}a, while the experimental systematics breakdown is
summarized in Table~\ref{tab:systbreak}. At \ycut~values above 0.004 the
measurement is dominated by statistical uncertainties, while for very low
values of \ycut~the data samples increase and the global normalisation
uncertainty dominates. Gluon splitting uncertainties are kept low in the whole
\ycut~range thanks to the dedicated anti-gluon splitting cut 
(see Section~\ref{sec:fourjettag}). 
\Herwig~provides the best description, being compatible with the
experimental data in the whole \ycut~range. However, the \Pythia~prediction is
only 1.5 standard deviations away in the large \ycut~region;
\Ariadne~provides a good
description of the data in the region $\ycut\geq 0.005$, while for lower
values of \ycut~it tends to underestimate the mass effect. 

\subsection{Experimental uncertainties}
\label{sec:expsyst}
Experimental uncertainties arise in the process of correcting the detector-level
measurement to hadron level, and are due to imperfections in the physics and 
detector modelling in the \delsim~simulation used in the correction procedure. 
The following sources have been considered in this analysis: 
\bi
\item[$\bullet$] {\em Statistical}: these uncertainties are due to the limited
  size of the experimental and simulated data samples. They are estimated from
  a toy simulation based on Poisson statistics. Central values were taken from
  the data and simulated samples, and correlations between the different
  quantities were accounted for by building up the corresponding covariance
  matrix. 

\item[$\bullet$] {\em Gluon splitting}: the identification of primary $b$-quarks is based 
on the presence of long-lived $B$ and $D$-hadrons in the final state. However, light-quark
events with gluon radiation splitting into secondary heavy quarks can produce a similar
signature. The correction procedure is very sensitive to the gluon splitting rates in 
the Monte Carlo simulation through the signal and background efficiencies~\cite{MJ}. 
Their value was varied in the range of their quoted uncertainties~\cite{gsplit}
and the observed change in the observables was added in quadrature and taken to 
represent the corresponding uncertainty.

\item[$\bullet$] {\em Normalisation}: the uncertainty on the global normalisation 
$R_b/R_{\ell}$ is estimated by varying the world average values of $R_b$ and $R_{\ell}=(1-R_b-R_c)$ 
in the range of their quoted uncertainties~\cite{pdg}, and taking 
the maximum variation in the final observable as the global normalisation 
uncertainty. This results in a $6\tcperthousand$ relative uncertainty and 
is \ycut~independent. The uncertainty from the charm-/light-quark normalisation
factor ($R_n^{c\ell}=R_n^c/R_n^{\ell}$) is estimated as half the maximum difference obtained 
by using as input to the measurement the prediction from the three 
event generators used: \Pythiav, \Herwigv~and \Ariadnev. 

\item[$\bullet$] {\em Flavour-tagging}: signal efficiencies 
($\epsilon_{nB}^b$ and $\epsilon_{nL}^{udsc}$) 
are measured from data and therefore do not
contribute to the total uncertainty for the double-tag technique.
To estimate the uncertainty due to the imperfect description of background
efficiencies and flavour correlations ($\rho_{nQ}^q$) in the simulation, the
calibration of the $b$-tagging in the simulation was exchanged with the
calibration obtained from data, which gives a poorer description of the 
lifetime probability~\cite{delphiRb}. 
Twice the observed difference was conservatively taken as the
flavour-tagging uncertainty. For the event-tag technique, the related
uncertainty was estimated as in~\cite{MJ} by varying the tagging
efficiencies within their uncertainties:
$\Delta\epsilon_{nB}^b/\epsilon_{nB}^b=3\%$ and
$\Delta\epsilon_{nL}^\ell/\epsilon_{nL}^\ell=8\%$ evaluated 
in reference~\cite{delphiRb}. 
The effect of mistagging efficiency was estimated by considering light-tagging as
equivalent to anti \qb-tagging,
i.e. $\Delta\epsilon_{n\ell}^q=\Delta\epsilon_{nb}^q$ for $q=b,c,\ell$ for the
same cut value.  
\ei

\subsection{Hadronisation corrections}
%
To compare parton-level fixed order ME calculations of $R_4^{b\ell-part}$ 
with experimental results, they must be corrected for hadronisation effects:
\beq
   R_4^{b\ell} = H_4^{b\ell} R_4^{b\ell-part} .
\eeq
The corrections $H_4^{b\ell}(\ycut)$ relating parton to hadron observables are
taken to be linear bin-to-bin factors.  

Three different generators, each tuned independently to the \delphi~data 
\cite{delphituning}, were used in this analysis: \Pythiav, \Herwigv~and
\Ariadnev. It was found that the \Herwig~and
\Ariadne~event generators are consistent both with the theoretical predictions
at the parton level (within the theoretical uncertainty) and the data (see
Figure~\ref{fig:4j_generators}) for a large range of \ycut. The hadronisation
corrections computed with the three generators are shown in
Figure~\ref{fig:4j_generators}b. 
The average of the \Herwig~and \Ariadne~predictions was used 
to correct the massive ME theoretical calculations 
(in the region of \ycut~studied here,
the hadronisation correction computed from \Pythia~is contained in the band
defined by the \Herwig~and \Ariadne~corrections).


\subsection{$b$-quark mass extraction and approximate NLO ME calculation}
\label{sec:4jettheo}
For a given flavour $q$, the $n$-jet rate is defined as the normalised $n$-jet 
cross-section
$R_n^q=[\Gamma_n/\Gamma_{tot}]^{Z\ra\qq}$. Theoretically, it
is convenient to use the double-ratios $\Rnbl=R_n^b/R_n^\ell$ as in this
observable most of the higher-order electroweak corrections, the first order
dependence on \alphas~and, to some extent also neglected higher-order terms in
\alphas, cancel out. Massive ME theoretical
calculations exist up to order $\alphas^2$~\cite{nlo1,nlo2,nlo3} and describe 
the 2, 3 and 4-jet rates for heavy ($b$, $c$) and light quarks ($\ell = uds$). 
Such calculations, when performed
in the $on-shell$ scheme in terms of the pole mass, $M_q$, can be
rewritten in terms of the running mass, $m_q$, defined in the $\MS$ scheme,
using the following order \alphas~relation: 
\beq
\label{eq:pole2run}
     M_q^2 = m_q^2(\mu) 
             \left[ 1 + \frac{\alphas}{\pi}
                        \left( \frac{8}{3}-2\log\frac{m_q^2(\mu)}{\mu^2} \right)
                      +{\mathcal{O}}(\alphas^2) \right].
\eeq
\noindent
Both mass definitions are equivalent at LO (see Eq. \ref{eq:pole2run}). 
For $\ycut=0.0065$, a value within a region with good stability, high
sensitivity and small hadronisation corrections,
the following $b$-quark mass value was obtained: 
$$ M_b=m_b(M_Z)=3.76 \pm 0.32~({\rm stat}) \pm 0.17~({\rm syst}) 
                     \pm 0.22~({\rm had}) \pm 0.90~({\rm theo})~\GeVcTwo . $$
The theoretical uncertainty is estimated as half the difference between the
$\Rfbl$ LO prediction for the running and pole $b$-quark mass definitions 
(see Figure~\ref{fig:4j_generators}b).

\begin{figure}[!htb]
\centering
  \includegraphics[width=0.495\linewidth]
  {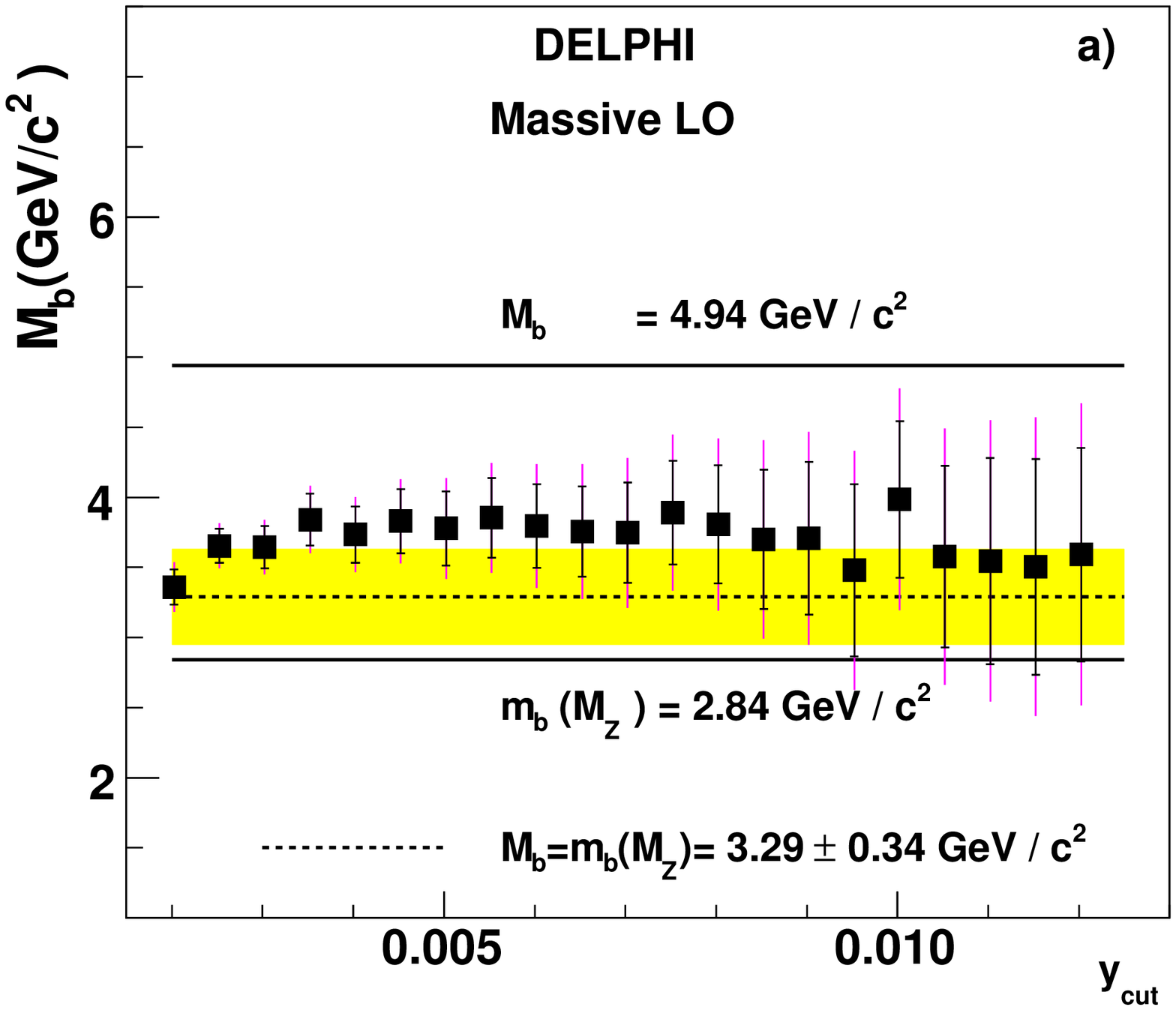}\\ 
  \includegraphics[width=0.495\linewidth]
  {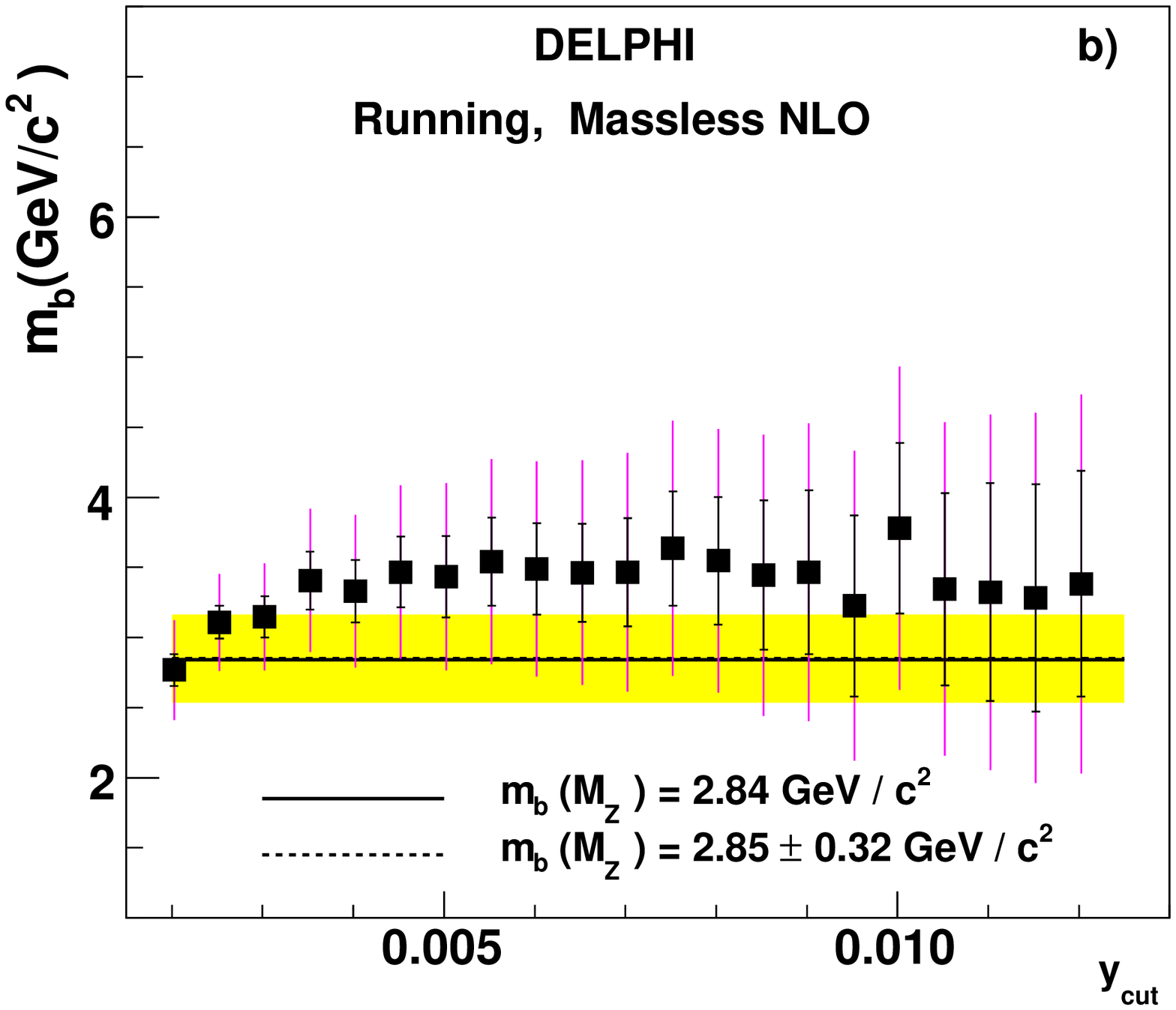}
  \includegraphics[width=0.495\linewidth]
  {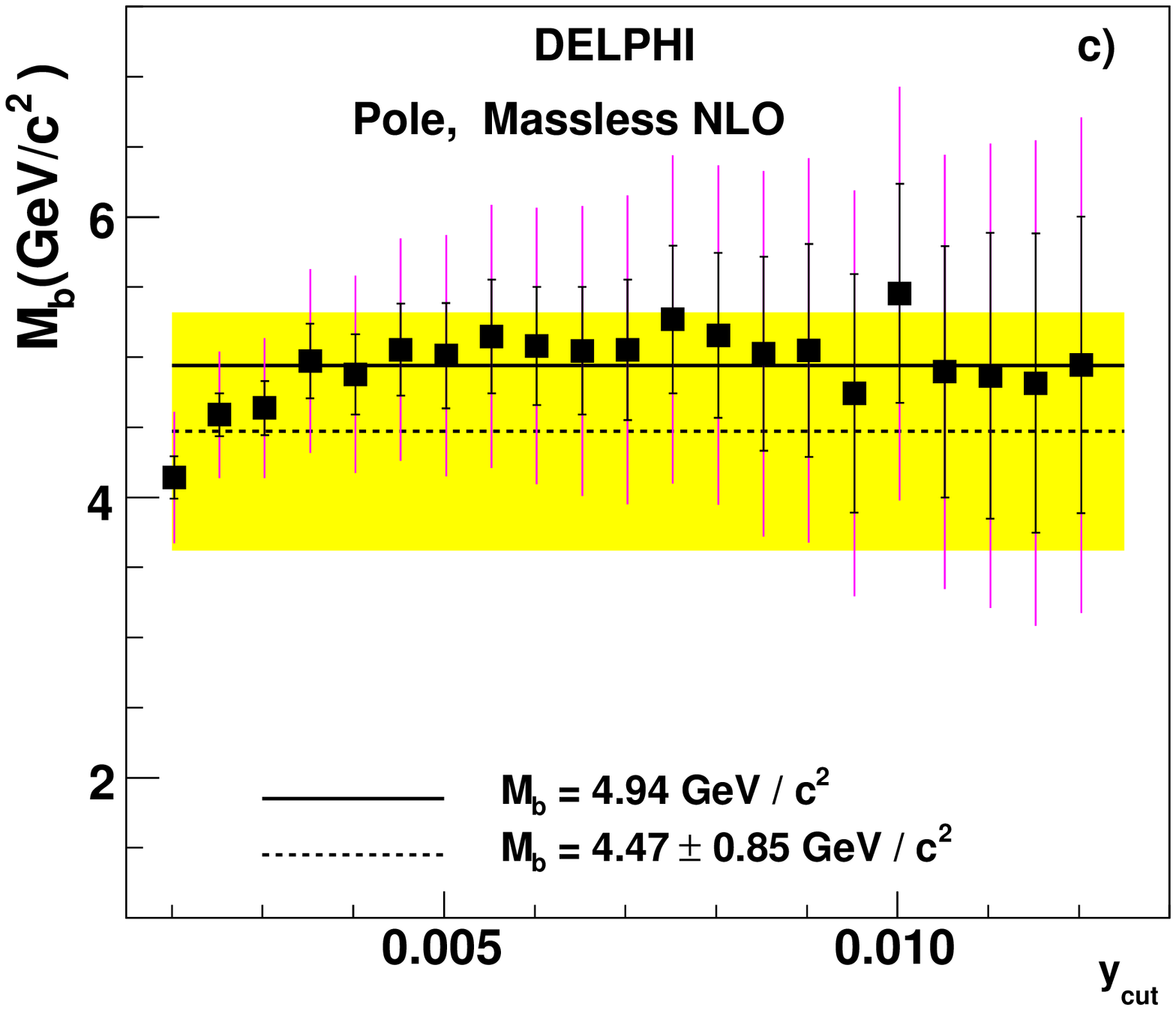}
 \caption{ (a) Massive LO results extracted from \Rfbl~(data points)
 compared with the  
 result obtained at LO in the \Rthreebl~analysis~\cite{MJ}: 
 $\Mb=\mb(M_Z)=3.29 \pm 0.34\,\GeVcTwo$. 
 In the LO result, no theoretical uncertainties are shown. 
 Results obtained from \Rfbl~using 
 massless NLO corrections include theoretical uncertainties estimated as 
 explained in Section~\ref{sec:4jettheo}. They are shown for the (b) running 
 and (c) pole mass definitions and are compared with the results obtained at
 NLO in the \Rthreebl~analysis~\cite{MJ}:  $\mb(M_Z)=2.85 \pm 0.32\,\GeVcTwo$ 
 and $\Mb=4.47 \pm 0.85\,\GeVcTwo$, respectively, shown as $\pm 1 \sigma$ 
 shaded band with its central value as a dotted line. Predicted values from the QCD 
 calculations at low energy, described in Section~\ref{sec:4jettheo}, 
 are also shown as solid lines. 
 } 
\label{fig:4jmasstest}
\end{figure}

To extract a meaningful $b$-quark running mass from the four-jet observable by
means of Eq.~\ref{eq:pole2run}, the NLO correction to $R_n^q$ would be needed,
which is only available for massless quarks~\cite{debrecen1,debrecen2}.  
However, an improvement of the LO estimation can be obtained if most of the
mass effect is contained in the LO term and hence the NLO correction
to \Rfbl~can be approximated as massless~\cite{private_german}:     
\beq
\label{eq:4jnlo}
 \Rfbl=\frac{A^b(m_b)\alphas^2+B^\ell\alphas^3}{A^\ell\alphas^2+B^\ell\alphas^3}, 
\eeq 
where the LO functions $A^b,A^\ell$ are taken from~\cite{nlo1,nlo2,nlo3} 
and the NLO massless term $B^\ell$ from~\cite{debrecen1,debrecen2}. 
As for the case of \Rthreebl~\cite{cite9}, it was found that:
\bi
\item[$\bullet$] the NLO corrections using the pole and running mass definitions
 were both within the uncertainty band defined by the two LO curves;
\item[$\bullet$] the running mass definition results in a smaller correction 
 at NLO than the pole mass.
\ei
The $b$-mass values obtained from \Rfbl~using this approximation are shown 
in Figure~\ref{fig:4jmasstest}b-c. They are found to be stable in the region
$\ycut>0.003$ and consistent with mass results obtained from $\Rthreebl$
(both at LO and NLO) and predicted values from QCD calculations at low energy
evolved to $M_Z$ using the RGE. 
For the running mass calculation, the massless NLO 
correction is small and results in very little effect. On the contrary,
for the pole mass the NLO correction is about $10\%$, 
leading to sizeable effects.

For the running $b$-quark mass definition, the theoretical prediction of
$\Rfbl$ is taken to be the central value of the following, 
in principle equivalent, four calculations: 
(a)~Full ratio as in Eq.~\ref{eq:4jnlo}, expressed in terms of
the running mass by means of Eq.~\ref{eq:pole2run} at the scale $\mu=\MZ$; 
(b)~Same, but using Eq.~\ref{eq:pole2run} at an arbitrary scale $\mu_0=M_b$ and
evolving the result to $\mu=\MZ$ via the RGE to obtain $\mb(\MZ)$; 
(c)~Series expansion of Eq.~\ref{eq:4jnlo}, 
expressed in terms of $\mb(\MZ)$ as in the first method; 
(d)~Same, but introducing an arbitrary intermediate scale as in
the second method. The pole mass prediction is obtained in a similar way. The
resulting predictions for \Rfbl~are shown in Figure~\ref{fig:4j_generators}b
for a reference \qb-quark mass obtained by evolving the average of low energy
measurements $\mb(\mb)=4.20 \pm0.07~\GeVcTwo$~\cite{pdg}  
to the $M_Z$ scale, $\mb(\MZ)=2.84\pm 0.06~\GeVcTwo$, or by translating it to
a pole mass value: $\Mb = 4.94 \pm 0.08\,\GeVcTwo$. The strong coupling
constant value used was $\alphas(M_Z)=0.1202\pm 0.0050$~\cite{worldalphas}. 

\subsection{Theoretical and modelling uncertainties}
\label{sec:theosyst}
The following sources of systematic uncertainty have been considered for the 
comparison
of the corrected four-jet ME calculations with the experimental results:
\bi
\item[$\bullet$] {\em Theoretical uncertainties}, due to missing higher orders 
  in matrix element calculations and to the use of massless next-to-leading 
  corrections for
  the mass extraction, cannot be rigorously estimated in the case of four-jets. 
  However, following a comparison
  between the same approximation applied to $\Rthreebl$ with the full massive
  calculation available in this case, this uncertainty 
  was conservatively taken to be twice the maximum difference between the 
  four predictions defined in Section~\ref{sec:4jettheo}. 
  The theoretical uncertainty is responsible for about $0.4-0.5~\GeVcTwo$ 
  in the uncertainty of the final result, and it is almost independent of
  \ycut. Although lower than in the case of the LO calculation, it is three
  times higher than in the completely massive three-jet calculation.

\item[$\bullet$] {\em Modelling uncertainties}, related to the correction for
  hadronisation effects of the theoretical calculations at parton level using 
  Monte Carlo event generators. 
  This includes the uncertainty on the tuned values of the free parameters
  in each model (including the $b$-mass parameter entering in the parton 
  shower~\cite{MJ}) and the modelling of hadronisation. 
  The size of the modelling uncertainty is estimated as half the difference 
  between the predictions from \Herwigv~and \Ariadnev\footnote{The result obtained with the
  \Pythiav~event-generator is compatible with the quoted results within the 
  modelling uncertainty.}. 
  To include the $b$-mass uncertainty in the estimation of the hadronisation 
  systematic uncertainty, the mass parameter in \Herwig~and \Ariadne~was varied 
  within $\pm 0.125~\GeVcTwo$ around their central values in order to maximize 
  the difference 
  between both predictions. This was achieved by setting the mass parameter to 
  $M_b=4.85~\GeVcTwo$ in both generators. 
  The total modelling uncertainty amounts to $\pm(1-2)\%$
  in the region of $\ycut>0.004$, corresponding to about $\pm 0.2~\GeVcTwo$ 
  in terms of both the running and pole mass results. The contribution from
  varying the mass parameter amounts to about $\pm 0.1~\GeVcTwo$.
\ei
The breakdown of the theoretical and modelling uncertainties in the
$b$-quark mass results obtained from $R_4^{b\ell}$ is detailed 
in~Table~\ref{tab:systbreak}.

\section{Summary and conclusions}
\label{sec:summary}
A new determination of the hadron-level $R_n^b$ and $R_n^{\ell=uds}$ jet-rates
($n=2,3,4$ jets) has been performed, using flavour tagging only in each
$n$-jet sample and obtaining the global normalisation of the observables from
the world average \Rb~and \Rc~measurements~\cite{pdg}. This measurement is
based on a double-tag technique which measures the flavour-tagging 
efficiencies directly from data, thereby reducing systematic uncertainties. 

Double-ratio observables are also studied: \Rfbl~is obtained from the four-jet 
rates $R_4^b$ and $R_4^\ell$ using this double-tag technique, and \Rtwobl~using
the event-tag method defined in reference~\cite{MJ}. 
Results from \Rtwobl~(and from the
previous measurements of \Rthreebl~in~\cite{MJ}) are also cross-checked.

Results are presented at hadron level, in order to allow for future
comparisons without having to unfold hadronisation and detector corrections
applied to the data (a summary of jet-rate results as a function of \ycut~is
shown in Tables~\ref{tab:systbreakall}-\ref{tab:systbreakalltwo}). 
They are compared to three Monte Carlo event generators:
\Pythiav, \Herwigv~and \Ariadnev, tuned to \delphi~data~\cite{delphituning}. 
The \Herwigv~generator gives the best overall description of 
flavour jet-rates, $R_n^b$ and $R_n^\ell$, 
but \Ariadnev~provides the best results 
for $R_n^\ell$.
For double-ratios, \Herwigv~gives
also the best description. However, the two-jet observable \Rtwobl~is not
satisfactorily described by any of the three generators considered. 

\begin{figure}[hbtp]
\centering
  \epsfig{file=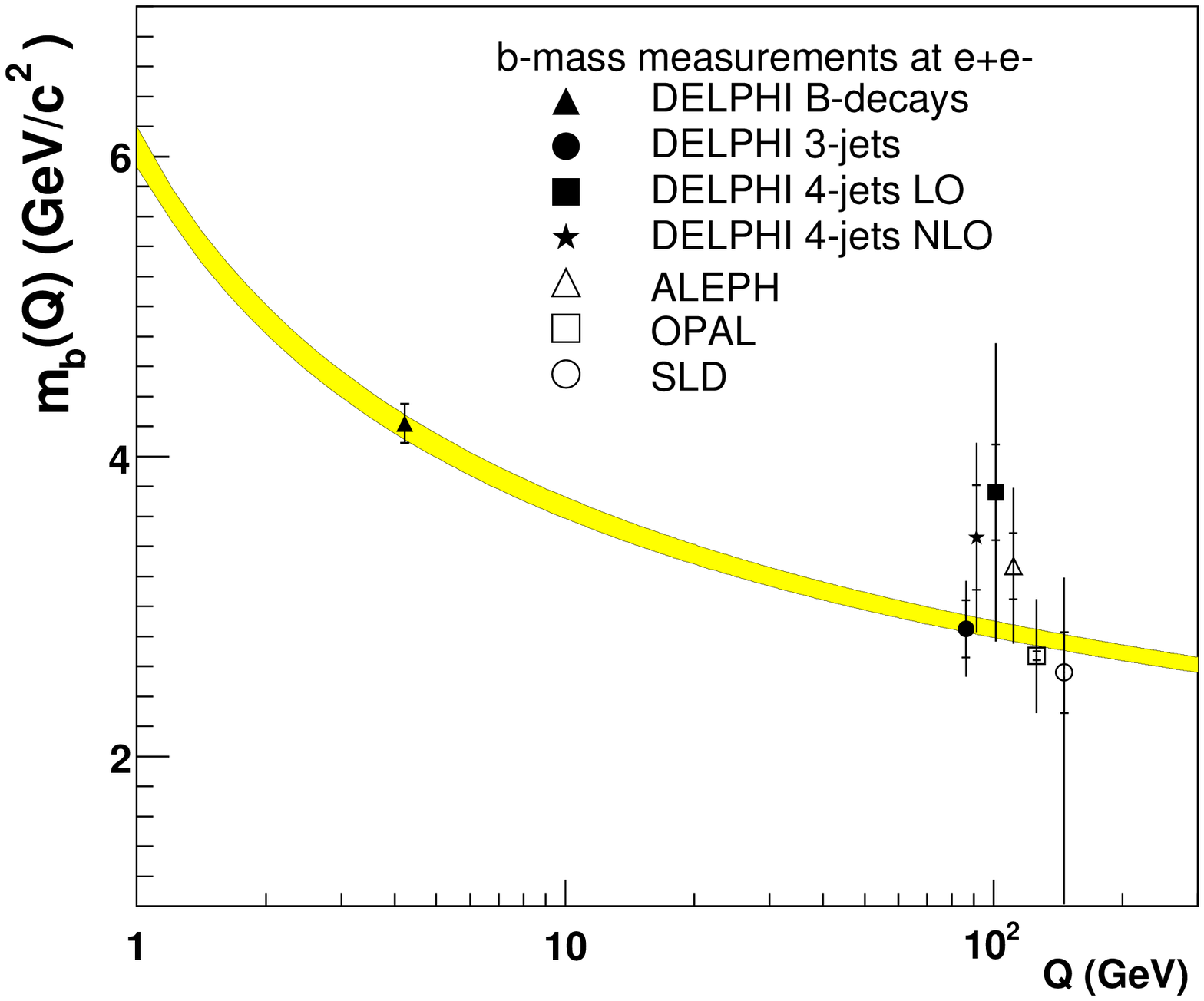, width=1\linewidth}
 \caption{The energy evolution of the \MS-running \qb-quark mass $\mb(Q)$
        as measured at \lep. \delphi~results from~\Rthreebl~\cite{MJ} at the
        $M_Z$ scale and from semileptonic $B$-decays~\cite{arantzamb} at low
        energy are shown together with results from other experiments
        (\daleph~\cite{alephmb}, \dopal~\cite{opalmb} and \sld~\cite{sldmb}).
        The masses extracted from LO and approximate NLO calculations of \Rfbl~are
        found to be consistent with previous
        experimental results and with the reference value $\mb(Q)$ (grey
        band) obtained from evolving the average $\mb(\mb)=4.20 \pm
        0.07~\GeVcTwo$ from~\cite{pdg} using QCD RGE (with a strong coupling constant value
        $\alphas(M_Z)=0.1202\pm 0.0050$~\cite{worldalphas}).} 
\label{fig:mbmz_delphi}
\end{figure}

A new determination of the $b$-quark mass in the four-jet topology has been
performed using the \cambridge~jet-clustering algorithm~\cite{cambridge}. 
The mass is measured
by comparing the experimental results of \Rfbl~at $\ycut=0.0065$ with fixed
order ME massive LO calculations assuming the universality of the strong
coupling constant, \alphas. The measured value is:
$$ m_b(M_Z) = 3.76 \pm 0.32~({\rm stat}) \pm 0.17 ~({\rm syst}) \pm 0.22~({\rm had}) 
                   \pm 0.90~({\rm theo})~\GeVcTwo . $$ 
A procedure to approximate the NLO corrections with the massless component in
order to improve the result has been tested successfully with the three-jet
massive calculations. The measured value of the running \qb-quark mass when
applying this method to the four-jet observable is:
$$ \mb(\MZ) = 3.46 \pm 0.35~({\rm stat}) \pm 0.19~({\rm syst}) \pm 0.24~({\rm mod}) 
                   \pm 0.44~({\rm theo})~\GeVcTwo $$
and the corresponding value for the pole mass is:
$$ \Mb      = 5.07 \pm 0.46~({\rm stat}) \pm 0.24~({\rm syst}) \pm 0.32~({\rm mod}) 
                   \pm 0.57~({\rm theo})~\GeVcTwo. $$
These results agree within the uncertainties with the values obtained evolving the
average of low energy measurements $\mb(\mb)=4.20
\pm0.07~\GeVcTwo$~\cite{pdg} to the $M_Z$ scale using the RGE: $\mb(\MZ)=2.84\pm 0.06~\GeVcTwo$, or by translating it to a
pole mass value: $\Mb = 4.94 \pm 0.08\,\GeVcTwo$.
The values of $m_b(M_Z)$ obtained from the LO and approximate NLO $\Rfbl$~calculations are shown in 
Figure~\ref{fig:mbmz_delphi} together with results from other measurements at 
the $M_Z$ scale, in particular the most precise result from \Rthreebl,
$m_b(M_Z)=2.85\pm0.32\,\GeVcTwo$~\cite{MJ}, as well as results at low energy
from semileptonic $B$-decays~\cite{arantzamb} obtained at a lower mass scale. 
All experimental results are 
consistent with each other assuming the QCD running prediction from RGE. 

The main limitation in the extraction of $m_b(M_Z)$ from the \Rfbl~measurement 
is theoretical. 
If a calculation with resummed LL logarithms~\cite{massivenll,massivenll2}
could be used, a larger range of \ycut~could be exploited. This could
potentially lead to a lower uncertainty.

Improvements to the precision of $\mb(\MZ)$ are not expected from combining
the different measurements because they are largely limited by common systematic
uncertainties. Other methods will likely be needed at future colliders in
order to obtain more precise determinations of the \qb-quark mass at high energy. 
This will be important to interpret the precise measurements at the Linear
Collider in searches for new physics. As an example, a future linear collider
operating at $\sqrt{s}=500\,\GeV$ will produce Higgs bosons copiously (if they
exist). Since the decay branching fraction into \qb-quarks is expected to be
proportional to the mass squared, measurements of this decay channel would be
very sensitive to the exact value of the mass at that scale.  
\begin{table}[!hbtp]
\centering
\vspace{7mm}
\begin{tabular}{|c|c|c|c||c|c|}
\hline
\ycut   & $R_2^b$   & $R_2^\ell$ & \Rtwobl    &   $R_3^b$  & $R_3^\ell$\\
\hline
0.003   & $0.505 \pm 0.003$    & $0.481\pm 0.003$     & $1.062 \pm 0.004$ & $0.342 \pm 0.002$     & $0.355 \pm0.003$  \\
\hline
0.004   & $0.553 \pm0.003$     & $0.527\pm 0.003$     & $1.060 \pm 0.003$ & $0.329 \pm 0.002$     & $0.345 \pm0.003$    \\
\hline
0.005   & $0.585 \pm0.003$    & $0.562\pm 0.003$     & $1.053\pm 0.003$   & $0.317 \pm 0.002$     & $0.333 \pm 0.003$    \\
\hline
0.006   & $0.611 \pm0.003$     & $0.591\pm 0.003$    & $1.046\pm 0.003$ & $0.306 \pm 0.003$      & $0.321 \pm0.003$    \\
\hline
0.007   & $0.633 \pm0.003$     & $0.615 \pm 0.003$     & $1.040\pm 0.003$ & $0.295 \pm 0.003$      & $0.310 \pm0.003$    \\
\hline
0.008   & $0.651\pm0.003$     & $0.635 \pm 0.003$     & $1.040 \pm 0.003$ & $0.286 \pm 0.002$ & $0.300 \pm0.003$    \\
\hline
0.009   & $0.667\pm0.003$     & $0.653 \pm 0.003$    & $1.031\pm 0.003$   & $0.277 \pm 0.002$     & $0.289 \pm0.003$    \\
\hline
0.010   & $0.681\pm0.003$     & $0.669\pm 0.004$     & $1.027\pm 0.002$       & $0.268 \pm 0.002$     & $0.280\pm0.003$    \\
\hline
\end{tabular}
\caption{Summary of experimental two and three-jet rates, with their total uncertainty,
         as a function of \ycut~\cite{cambridge}.}
\label{tab:systbreakall}
\end{table}

\begin{table}[!hbtp]
\centering
\vspace{7mm}
\begin{tabular}{|c|c|c|c|}
\hline
\ycut   & $R_4^b$    & $R_4^\ell$  & \Rfbl \\
\hline
0.003   & $0.1148\pm 0.0013$      & $0.1248\pm 0.0009$      & $0.920\pm 0.013$  \\
\hline
0.004   & $0.0911\pm 0.0012$      & $0.1018\pm 0.0007$      & $0.895\pm 0.015$ \\
\hline
0.005   & $0.0757\pm 0.0012$      & $0.0856\pm 0.0007$      & $0.885\pm 0.016$ \\
\hline
0.006   & $0.0642 \pm 0.0011$     & $0.0729\pm 0.0007$      & $0.882\pm 0.017$ \\
\hline
0.007   & $0.0555\pm 0.0010$      & $0.0628\pm 0.0006$      & $0.884\pm 0.019$ \\
\hline
0.008   & $0.0486\pm 0.0010$      & $0.0551\pm 0.0006$      & $0.88\pm 0.03$  \\
\hline
0.009   & $0.0432\pm 0.0010$      & $0.0486\pm 0.0005$      & $0.89\pm 0.02 $  \\
\hline
0.010   & $0.0380\pm 0.0010$      & $0.0431\pm 0.0005$      & $0.88\pm 0.03 $  \\
\hline
\end{tabular}
\caption{Summary of experimental four-jet rates, with their total uncertainty,
         as a function of \ycut~\cite{cambridge}.}
\label{tab:systbreakalltwo}
\end{table}

\newpage
\section*{Acknowledgements}
We are grateful to G. Rodrigo and A. Santamar\'{\i}a for providing
the theoretical input for this measurement. We are also indebted to 
T. Sj\"ostrand for his help in understanding how mass effects are implemented
in \Pythia. We would also like to thank G. Dissertori for continuous
advice and J. Portoles and M. Eidem\"uller for their information about the
$b$ pole mass. 

We are greatly indebted to our technical 
collaborators, to the members of the CERN-SL Division for the excellent 
performance of the LEP collider, and to the funding agencies for their
support in building and operating the DELPHI detector.
We acknowledge in particular the support of \\
Austrian Federal Ministry of Education, Science and Culture,
GZ 616.364/2-III/2a/98, \\
FNRS--FWO, Flanders Institute to encourage scientific and technological
research in the industry (IWT) and Belgian Federal Office for Scientific,
Technical and Cultural affairs (OSTC), Belgium, \\
FINEP, CNPq, CAPES, FUJB and FAPERJ, Brazil, \\
Ministry of Education of the Czech Republic, project LC527, \\
Academy of Sciences of the Czech Republic, project AV0Z10100502, \\
Commission of the European Communities (DG XII), \\
Direction des Sciences de la Mati$\grave{\mbox{\rm e}}$re, CEA, France, \\
Bundesministerium f$\ddot{\mbox{\rm u}}$r Bildung, Wissenschaft, Forschung 
und Technologie, Germany,\\
General Secretariat for Research and Technology, Greece, \\
National Science Foundation (NWO) and Foundation for Research on Matter (FOM),
The Netherlands, \\
Norwegian Research Council,  \\
State Committee for Scientific Research, Poland, SPUB-M/CERN/PO3/DZ296/2000,
SPUB-M/CERN/PO3/DZ297/2000, 2P03B 104 19 and 2P03B 69 23(2002-2004),\\
FCT - Funda\c{c}\~ao para a Ci\^encia e Tecnologia, Portugal, \\
Vedecka grantova agentura MS SR, Slovakia, Nr. 95/5195/134, \\
Ministry of Science and Technology of the Republic of Slovenia, \\
CICYT, Spain, AEN99-0950, AEN99-0761 and IN2P3/CYCIT bilateral funding agreement PP01/1,  \\
The Swedish Research Council,      \\
Particle Physics and Astronomy Research Council, UK, \\
Department of Energy, USA, DE-FG02-01ER41155, \\
EEC RTN contracts HPRN-CT-00292-2002 and RTN2-2001-00450.

\newpage


\newpage
\begin{thebibliography}{ref99}

\bibitem{massycut}
G. Rodrigo, M. Bilenky, A. Santamar\'{\i}a, Nucl. Phys. {\bf B439} (1995) 505.
\bibitem{delmbmz}
DELPHI Coll., P. Abreu et al., Phys. Lett. {\bf B418} (1998) 430; \\
S. Mart\'{\i} i Garc\'{\i}a, J. Fuster and S. Cabrera, Nucl. Phys. {\bf B} 
Proc. Suppl. {\bf 64} (1998) 376.
\bibitem{alephmb}
ALEPH Coll., R. Barate \etal., Eur. Phys. J. {\bf C18} (2000) 1.
\bibitem{opalmb}
OPAL Coll., G. Abbiendi \etal., Eur. Phys. J. {\bf C21} (2001) 411.
\bibitem{sldmb}
A. Brandenburg \etal., Phys. Lett. {\bf B468} (1999) 168. 
\bibitem{MJ}
DELPHI Coll., J. Abdallah et al., Eur. Phys. J. {\bf C46} (2006) 569.
\bibitem{cambridge} 
Yu.L. Dokshitzer \etal., JHEP {\bf 9708} (1997) 001.
\bibitem{pythia}
T. Sj$\ddot{{\mathrm o}}$strand \etal., Comp. Phys. Comm. {\bf 135} (2001) 238; \\
T. Sj$\ddot{{\mathrm o}}$strand \etal., {\em PYTHIA 6.2 Physics and Manual}, hep-ph$/$0108264.
\bibitem{herwig}
G. Marchesini \etal., Comp. Phys. Comm. {\bf 67} (1992) 465; \\
G. Corcella \etal., JHEP {\bf 0101} (2001) 010.
\bibitem{ariadne}
L. L$\ddot{{\mathrm o}}$nnblad, Comp. Phys. Comm. {\bf 71} (1992) 15.
\bibitem{nlo1} 
G. Rodrigo, M. Bilenky, A. Santamar\'{\i}a, Phys. Rev. Lett. {\bf 79} (1997) 193.
\bibitem{nlo2} 
W. Bernreuther, A. Brandenburg, P. Uwer, Phys. Rev. Lett. {\bf 79} (1997) 189.
\bibitem{nlo3}
P. Nason and   C. Oleari, Phys. Lett. {\bf B407} (1997) 57.
\bibitem{pdg}
Particle Data Group, W.-M. Yao \etal., J. Phys. {\bf G33} (2006) 1.
\bibitem{debrecen1}
Z. Nagy and Z. Trocsanyi, Phys. Rev. Lett. {\bf 79} (1997) 3604.
\bibitem{debrecen2}
Z. Nagy and Z. Trocsanyi, Phys. Rev. {\bf D59} (1999) 014020; Erratum-ibid.
{\bf D62} (2000) 099902. 
\bibitem{private_german}
J. Drees and G. Rodrigo, private communication.
%
%
\bibitem{delphi1}
DELPHI Coll., P. Aarnio \etal., Nucl. Instr. and Meth. {\bf A303} (1991) 233.
\bibitem{delphi2}
DELPHI Coll., P. Abreu \etal., Nucl. Instr. and Meth. {\bf A378} (1996) 57.
\bibitem{delphituning}
DELPHI Coll., P. Abreu et al., Z. Phys. {\bf C73} (1996) 11;
for the tuning of \Herwigv, see~\cite{MJ}.
\bibitem{gsplit}
The LEP/SLD Heavy Flavour Working group, LEPHF/2001-01,
http://lepewwg.web.cern.ch/LEPEWWG/heavy/lephf0101.ps.gz.
%
%
\bibitem{newbtag}
DELPHI Coll., Eur. Phys. J. {\bf C32} (2004) 185.
\bibitem{delphiRb}
DELPHI Coll., Eur Phys. J. {\bf C10} (1999) 415.
%
%
%
%
\bibitem{cite9} 
M. Bilenky \etal., Phys. Rev. {\bf D60} (1999) 114006.
\bibitem{worldalphas} 
The LEP QCD Working Group, paper in preparation. 
\bibitem{arantzamb}
M. Battaglia \etal., Phys. Lett. {\bf B556} (2003) 41.
\bibitem{massivenll} 
F. Krauss and G. Rodrigo, Phys. Lett. {\bf B576} (2003) 135.
\bibitem{massivenll2}
S. Catani \etal., Nucl. Phys {\bf B627}
(2002) 189.
\end{thebibliography}
\end{document}